\documentclass[utf8]{frontiersFPHY} % for Physics and Applied Mathematics and Statistics articles
\setcitestyle{square} % for Physics and Applied Mathematics and Statistics articles
\usepackage{url,hyperref,lineno,microtype,subcaption} 
\usepackage[onehalfspacing]{setspace}

\def\firstAuthorLast{Salucci {et~al.}} %use et al only if is more than 1 author

\def\Authors{Paolo Salucci\, $^{1,*}$, Giampiero Esposito\,$^2$, Gaetano Lambiase\,$^3$, Emmanuele Battista\,$^{2,4,5}$, Micol Benetti \, $^{2,6}$, Donato Bini\,$^{2,7}$, Lumen Boco\, $^1$, Gauri Sharma\,$^1$, Valerio Bozza\,$^3$,
 Luca Buoninfante\,$^3$, Antonio Capolupo\, $^3$, Salvatore Capozziello\,$^{2,6}$, Giovanni Covone \,$^{2,6}$,
 Rocco D'Agostino\,$^{2,6}$, Mariafelicia DeLaurentis\,$^{2,6}$, Ivan De Martino\, $^{2,8}$, Giulia De Somma\,$^{2,6}$, Elisabetta Di Grezia\,$^2$, Chiara Di Paolo\,$^1$, Lorenzo Fatibene\,$^9$,
 Viviana Gammaldi\,$^{10}$, Andrea Geralico\, $^{2,7}$, Lorenzo Ingoglia\,$^2$, Andrea Lapi\, $^1$, Giuseppe G. Luciano\,$^3$, Leonardo Mastrototaro\,$^3$, Adele Naddeo\,$^2$, Lara Pantoni\,$^1$,
 Luciano Petruzziello\,$^{2,11}$, Ester Piedipalumbo\,$^{2,6}$, Silvia Pietroni\,$^3$, Aniello Quaranta\,$^3$, Paolo Rota\,$^3$, Giuseppe Sarracino\,$^2$, Francesco Sorge\, $^2$, Antonio Stabile\,$^3$, Cosimo Stornaiolo\,$^2$, Antonio Tedesco\,$^3$, Riccardo Valdarnini\,$^1$,
Stefano Viaggiu\,$^{2,12}$, Andy A. V. Yunge\,$^2$}
 
\def\Address{$^{1}$ SISSA, via Bonomea 262, Trieste, Italy \\ $^2$ INFN, Sezione di Napoli, C.U. Monte S. Angelo,
Via Cinthia, I-80126 Napoli, Italy \\ $^3$ Dipartimento di Fisica ``E.R. Caianiello'', Universita' degli Studi di Salerno, Via Giovanni Paolo II, 132, 84084 Fisciano (SA), Italy and INFN Gruppo Collegato di Salerno, Italy \\ 
$^{4}$ Institute for Theoretical Physics, Karlsruhe Institute of Technology, 76128 Karlsruhe, Germany. \\
$^{5}$ Institute for Nuclear Physics, Karlsruhe Institute of Technology, Hermann-von-Helmholtz-Platz 1, 76344 Eggenstein-Leopoldshafen, Germany \\
$^6$ Dipartimento di Fisica E. Pancini, Universita' di Napoli ``Federico II", Via Cinthia I-80126 Napoli \\ $^7$ Istituto per le Applicazioni del Calcolo "M. Picone," CNR via dei Taurini 19 I-00185 Roma (IT) \\ 
$^{8}$ Dipartimento di Fisica, University of Torino, Via P. Giuria 1, 10125 Torino \\
$^9$ Department of Mathematics Giuseppe Peano, Universita' degli Studi di Torino, via Carlo Alberto 10, I-10123 Torino, Italy and INFN Sezione di Torino, Via Pietro Giuria 1, I-10125 Torino, Italy.\\
 $^{10}$ Departamento de Fisica Teorica, Universidad Autonoma de Madrid, Madrid, Spain. Instituto de Fisica Teorica UAM-CSIC, Campus de Cantoblanco, Madrid, Spain. \\ 
 $^{11}$ Dipartimento di Ingegneria, Universita' di Salerno, Via Giovanni Paolo II, 132 I-84084 Fisciano (SA), Italy \\
$^{12}$ Dipartimento di Fisica Nucleare, Subnucleare e delle Radiazioni Universita' degli Studi Guglielmo Marconi, Via Plinio 44, I-00193 Roma, Italy 
}

\def\corrAuthor{Paolo Salucci}

\def\corrEmail{salucci@sissa.it}

\newcommand{\ltsima}{$\; \buildrel < \over \sim \;$}
\newcommand{\simlt}{\lower.5ex\hbox{\ltsima}}
\newcommand{\gtsima}{$\; \buildrel > \over \sim \;$}
\newcommand{\simgt}{\lower.5ex\hbox{\gtsima}}
\newcommand{\gr}{\kern 2pt\hbox{}^\circ{\kern -2pt K}} % ====> GRADI KELVIN

\begin{document}
\onecolumn
\firstpage{1}

\title[Einstein, Planck and Vera Rubin]{Einstein, Planck and Vera Rubin: relevant encounters between the Cosmological and the Quantum Worlds}

\author[\firstAuthorLast]{\Authors} %This field will be automatically populated
\address{} %This field will be automatically populated
\correspondance{} %This field will be automatically populated

\extraAuth{}%
%\extraAuth{corresponding Author2 \\ Laboratory X2, Institute X2, Department X2, Organization X2, Street X2, City X2 , State XX2 (only USA, Canada and Australia), Zip Code2, X2 Country X2, email2@uni2.edu}

\maketitle  
\begin{abstract}

 In Cosmology and in Fundamental Physics there is a crucial question like: where the elusive substance that we call Dark Matter is hidden in the Universe and what is it made of?, that, even after 40 years from the Vera Rubin seminal discovery  does not have a proper answer. Actually, the more we have investigated, the more this issue has become strongly entangled with aspects that go beyond the established Quantum Physics, the Standard Model of Elementary particles and the General Relativity and related to processes like the Inflation, the accelerated expansion of the Universe and High Energy Phenomena around compact objects. Even Quantum Gravity and very exotic DM particle candidates may play a role in framing the Dark Matter mystery that seems to be accomplice of new unknown Physics.
 
 Observations and experiments have clearly indicated that the above phenomenon cannot be considered as already theoretically framed, as hoped for decades. The Special Topic to which this review belongs wants to penetrate this newly realized mystery from different angles, including that of a contamination of different fields of Physics apparently unrelated. We show with the works of this ST that this contamination is able to guide us into the required new Physics. 
 
 This review wants to provide a good number of these "paths or contamination"
 beyond/among the three worlds above; in most of the cases, the results presented here open a direct link with the multi-scale dark matter phenomenon, enlightening some of its important aspects. Also in the remaining cases, possible interesting contacts emerges. Finally, a very complete and accurate bibliography is provided to help the reader in navigating all these issues. 
 
\end{abstract}

\section{Introduction}
 
\indent

The phenomenal roles in investigating the Universe and its content and governing rules, of these three great scientists is without discussion. However, despite that we have today a well organized system of laws of Nature and a formidable set of experimental and observational discoveries, the status of the Universe seems to require that researchers in Observational Cosmology, General Relativity and Quantum Physics concur together and join forces in building new effective paths of knowledge. 
In fact, it is well known that about $95\%$ of the energy density of the Universe is of dubious invisible nature and cannot be framed within the "presently verified" physics. Moreover, a number of phenomena like e.g. "inflation", "matter-antimatter asymmetry", "cosmological and astrophysical quantum relativistic objects", " dark matter" just to name a few, are far from being understood and good ideas like "the Unification of Gravity with the other forces" and "the intrinsic Symmetry of the Universe" are also stalling. 
 
In these circumstances, conceptually new directions of investigation have been proposed. The Universe is considered not only as the arena in which theories (e.g. on the Dark Matter) get falsified/verified, but, through the properties of its content, a strong motivator for creation of totally new ones. A contamination among Cosmology, Astrophysics, Relativity, Theoretical Physics, Physics of the Elementary particles theory is taking place in many recent investigations aimed at discovering new Physics. This is the leitmotiv of the Special Topic collection to which this work belongs; we want to contribute to it by providing a review of 17 examples of such new and in some case very different paths of contaminated knowledge. 

This review confirms that just the complementary of physical, astrophysical and cosmological probes leads to a successful approach to the (always more required) New Physics and Cosmology. This is done by describing a substantial number of topics of the above kind. Of course this work is not fully complete in reporting such investigations, so as also in exhaustively discussing all the most pressing arguments of Cosmology and Fundamental Physics. About this, let us stress, however, that this review is conceived in relation to the collection of articles of the Special Issue for which it works as the Introductory Paper. In addition, in order to widen the scope of the present work, we indicate to the reader the following reviews on crucial issues of Cosmology and Fundamental Physics which are related to those accurately dealt in the present work: "Cosmology Intertwined: Perspectives for the Next Decade"\cite{DiValentino:2020vhf, DiValentino:2020zio}, 
"Lectures on Black Holes and linear waves"\cite{Dafermos}, ``The distribution of dark matter in galaxies''\cite{sal20}, "Status of dark matter in the Universe" \cite{Freese_2017}, "Testing dark energy models in the light of $\sigma_8$ tension" \cite{Gaetanosigma8}. Finally, a large and accurate list of references will contribute to frame this review within the Cosmological and Fundamental Physics Literature.

\vskip 1.2cm 

\section{Cosmography in General Relativity and in Extended Theories of Gravity}
\label{Sec:Cosmography}
\vskip 0.6cm

 The standard cosmological model is also known as $\Lambda$CDM due to the dark components it takes on, i.e $\Lambda$ stands for the cosmological constant while CDM points to the Cold Dark Matter fluid. Such a model provides a successful description of the evolution and the structure of the universe, requiring just six parameters, but leaves several questions open including the nature of the dark matter and the cosmological constant. Recently, with the improvement of the accuracy of the cosmological observations, tensions between different probes have emerged. Among the most debated at the moment, there is certainly the tension in the Hubble constant measurements, $H_0$ when it is derived at early Universe, $z=1100$, from the CMB measurements and when it is measured at low redshift by exploiting the SNIa distance indicator calibrated by Cepheids. 
 Let us stress that, in the first case, the $H_0$ estimation strictly depends on the model used to describe the CMB data. In particular, the position of the main CMB peak reflects the size of the horizon at last scattering of the CMB photons and is determined almost entirely by the geometry of the Universe. The height of the peak is directly proportional to the fractional mass in baryons $\Omega_b$ and also varies according to the expansion rate of the Universe as specified by the Hubble constant $H_0$ \cite{Hancock:1996dq}. In the latter case it is instead an accurate estimate of the magnitude $\mu$ of the SNIa distance indicator calibrated by Cepheids, which implies a series of accurate techniques which also consider the Cepheid period, metallicity and luminosity (see Ref.\cite{Riess:2016jrr} for details). The magnitude is then related to the luminosity distance, $D_L$, via $ \mu= 5 log \, D_L + 25$, and it is possible to obtain the direct measurement (therefore independent of the model) of $H_0$ from the $D_L$ relationship, e.g $ D_L = (1 + z) \frac {zc}{H_0} $, that is valid for low redshift.

The tension on H0 mentioned above is quantified in about $4.4\sigma$ between the Planck collaboration estimation \cite{Aghanim:2018eyx} $H_0=\left( 67.27 \pm 0.60\right)$ km/s/Mpc and the SH0ES collaboration measurement \cite{Riess:2019cxk}) $H_0=(74.03\pm 1.42)$ km/s/Mpc using the Hubble Space Telescope observations of 70 long-period Cepheids in the Large Magellanic Cloud.

Furthermore, the existence of other mutually discrepant probes with no common observational systematics reinforces the idea that the above tension is systematics-free and due to unknown properties of the actual cosmological scenario \cite{Abbott:2017smn, Henning:2017nuy, Bonvin:2016crt}. As an example, we can cite a joint data analysis of Baryon Acoustic Oscillation (BAO) from BOSS and eBOSS and Big Bang Nucleosyntesis (BBN)~\cite{Alam:2020sor} that finds $H_0=(67.35 \pm 0.97)$, while SPTpol~\cite{Henning:2017nuy} finds $H_0=(71.3 \pm 2.1)$ km/s/M

Moreover, standard distance ladder and time delay distances agree with the SH0ES estimate. Finally, the H0LiCOW~\cite{Wong:2019kwg} value $H_0=(73.3^{+1.7}_{-1.8})$ km/s/Mpc, based on strong gravitational lensing effects on quasars favour high values for the Hubble constant as the result of the Maser Cosmology Project~\cite{Pesce:2020xfe}: $H_0=(73.9 \pm 3.0)$ km/s/Mpc.

These tensions may indicate an inadequacy in the Standard Cosmological model in providing a correct description of the evolution of the universe since the last scattering of the Cosmic Background Radiation till recent times. 

Extensions of the standard model, i.e. including {\it new physics} in the $\Lambda$CDM theory to obtain changes in observational predictions, were explored to solve the $H_0$ tension. Among the proposals, we can mention the inclusion of extra relativistic species at recombination \cite{Verde:2013cqa, Hou:2011ec, Trotta:2003xg, Graef:2018fzu, Benetti:2017juy} and the dynamical interaction between dark matter and dark energy \cite{Valentino1, Valentino2, Benetti:2019lxu}, as well as the investigation of the cosmological inflationary context \cite{Benetti:2019kgw}. Indeed, all these sectors can modify the value of the current expansion rate of the universe. For example, in the case of further relativistic species at CMB epoch, it is worth mentioning that the standard cosmological model considers three families of neutrinos as relativistic species at decoupling epoch. Nevertheless, also extra contributions can be considered as, for example, non-standard particle physics like eV-scale sterile neutrinos, Goldstone Bosons, thermal axions and even relativistic dark matter, or also the stochastic background of primordial gravitational waves \cite{cabass, meerburg}. These additional free-streaming relativistic particles are often called dark radiation~\cite{venninDR}. Intuitively, increasing the number of species increases the contribution of these species in the calculation of background evolution, thus increasing the value of the expansion rate. Despite the efforts spent to cover the shortcomings of the standard model by including new factors, the current conclusion is that such tension cannot be dealt within reasonable physics beyond the standard cosmological model since any resolution of this disagreement produces even greater tensions with other parameters of the model such as $\Omega_m$ and $\sigma_8$ \cite{ref1,ref2,ref3,ref4}.

The perspectives therefore open up, to explore drastically new theories, as the Extended Theories of Gravitation, or to discover the existence of subtle parameters' degeneracies plaguing the standard model. Cosmographic studies can give important hints also about the existing cosmological tensions (as those on $H_0$ or $\sigma_8$) since they are able to fully describe the expansion of the universe independently of any a-priori model, and therefore best highlight all the anomalies in the cosmological parameters that have not {\textit{a priori}} a known theoretical explanation \cite{Shafieloo11,rocco_review}.

\begin{figure}[h]
\begin{center}
\includegraphics[angle=0,width=.75\textwidth]{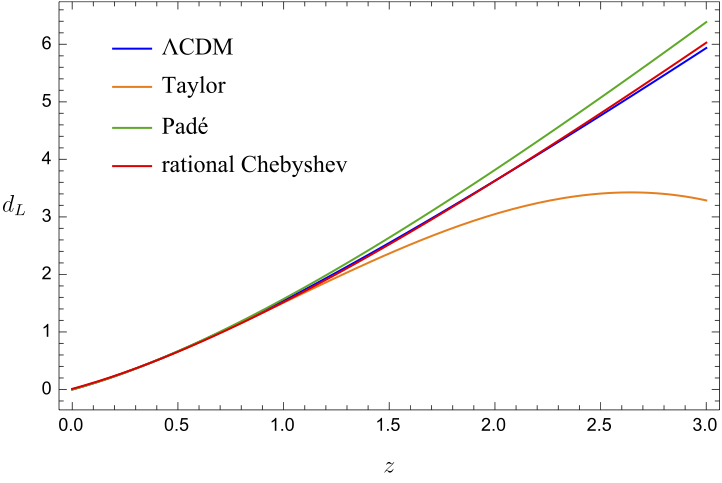}
\caption{Dimensionless luminosity distance curves for the $\Lambda$CDM model, the Taylor expansion at fourth-order, the Pad\'e(2,1) and Chebyshev(2,1) rational approximations.}
\label{fig_dagostino}
\end{center}
\end{figure}

The only assumptions cosmography relies on are the homogeneity and isotropy of the Universe. This model-independent technique permits the reconstruction of the dark energy dynamical evolution, with no need of assuming any particular cosmological model at late-time epochs \cite{Visser05,cosmogr20} (see Appendix~\ref{App:COSMOGRAPHY}).
Cosmography involves Taylor expansions of the observable quantities that get directly compared with data. The results of this procedure ensure the independence from a postulated equation of state for the Universe's evolution and, thus, help us to break the above degeneracy. However, this approach is plagued \cite{Dunsby16} by the divergence of the Taylor polynomials at high redshifts that leads to non-accurate numerical results when data at $z>1$ are used. To alleviate this problem one can use other polynomial expansions like the rational Pad\'e polynomials, which extend the convergence radius of the standard cosmographic series \cite{pade}, or the Chebyshev polynomials which reduce the systematics on the fitting coefficients and remain stable at high-redshifts. \cite{rocco_chebyshev} (cf. Fig. \ref{fig_dagostino}). 
  %%%%%%%%%%%%%%%%%%%%%%%%%%%%%%%%%%%%%%%%%%%%%%%%%%%%%%%%%%%%%
%%%%%%%%%%%%%%%%%%%%%%%%%%%%%%%%%%%%%%%%%%%%%%%%%%%%%%%%%%%%%%%%%%%
 At late times, assuming an arbitrary fiducial value for the current (dark + baryonic) matter density fraction, $\Omega_m$, the cosmography can be used to reconstruct the Hubble parameter in function of redshift using low redshift probes, as the Ia Supernovae. 
An extension of this strategy is to link together the parametric cosmographic behaviour of the late universe expansion with the small scale universe, i.e joining such a parametric late time evolution with a cosmological model as the $\Lambda$CDM (replacing the intake of dark energy with a cosmographic description of the current expansion of the universe). In principle, one can obtain a snapshot of the universe at any epoch. Following this strategy, it is possible to develop a $f(z)$CDM model, where $z$ is the redshift and the function $f(z)$ is a suitable combination of polynomials which track the cosmic luminosity distance and replace the cosmological constant $\Lambda$ at any $z$. This allows one to avoid the assumption of a fiducial value for $\Omega_m$, since the latter can be constrained by early universe data \cite{Benetti:2019gmo}. 
 %%%%%%%%%%%%%%%%%%%%%%%%%%%%%%%%%%%%%%%%%%%%%%%%%%%%
\begin{figure}[t]
\begin{center}
\includegraphics[width=0.85\textwidth]{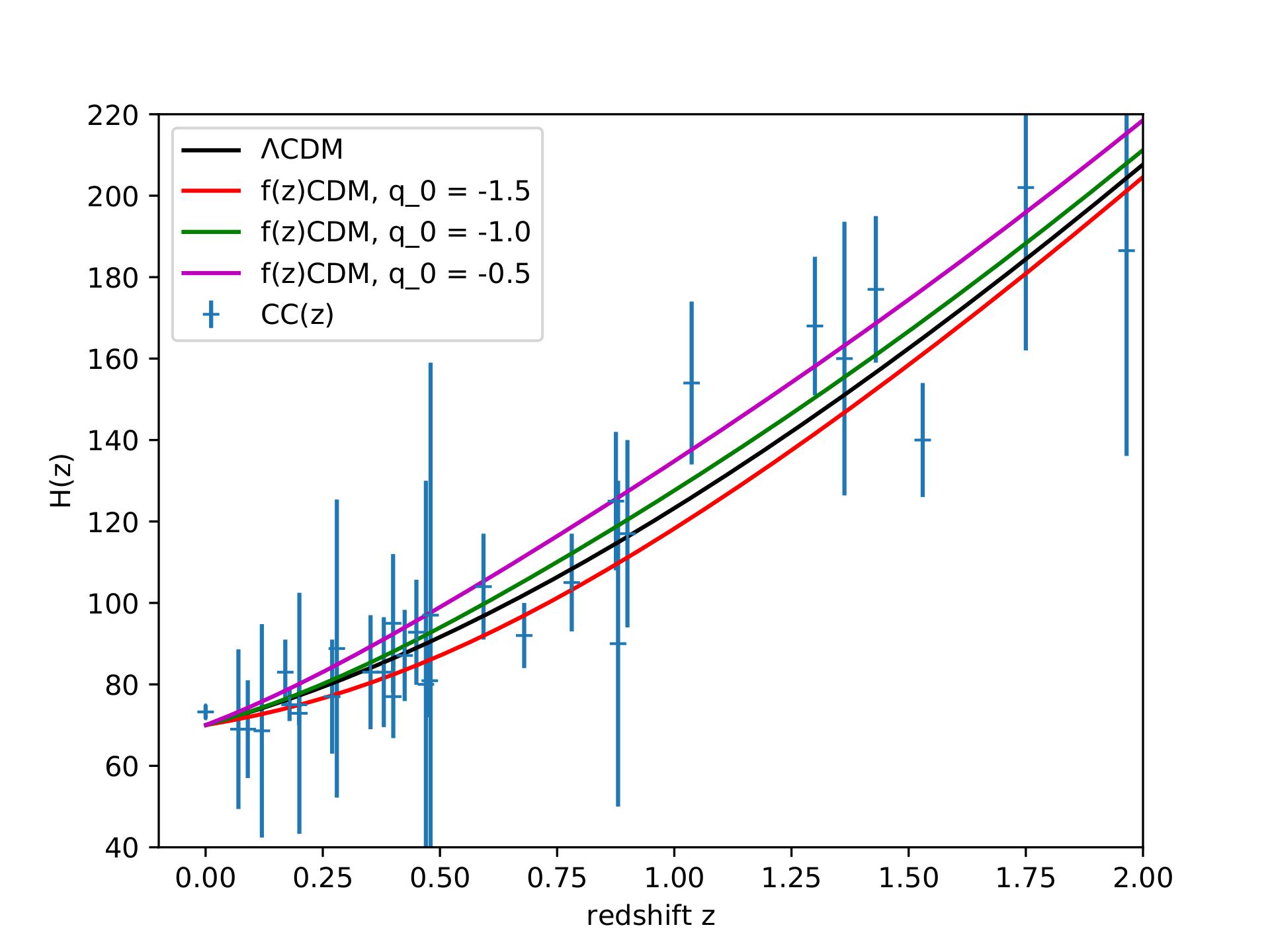}
\caption{$H(z)$ evolution expansion of $f(z)$CDM model, where $j_0=1.97$ and $s_0=l_0=0$ with different values of deceleration parameter: $q_0=-0.5$ (magenta curve), $q_0=-1$ (green curve), $q_0=-1.5$ (red curve). For comparison, the minimal $\Lambda$CDM model is also  drawn (black solid line) and the Cosmic Clock data are reported (see \cite{Benetti:2019gmo}). Moreover,  $\Omega_m = 0.3$, $\Omega_{\Lambda} = 1-\Omega_m$ and $H_0 = 70$ Km/s/Mpc.}
\label{fig_benetti}
\end{center}
\end{figure}
%%%%%%%%%%%%%%%%%%%%%%%%%%%%%%%%%%%%%%%%%%%%
The background evolution reads as: 

\begin{equation}
\label{eq:H_full}
E(z)^2 = \left(\frac{H(z)}{H_{0}}\right)^2 =  \Omega_r (1+z)^4 +  \Omega_m (1+z)^3 + \Omega_k (1+z)^2 +\Omega_f f(z)\,,
\end{equation}

with $\Omega_k$ the spatial curvature density, $\Omega_r$ the radiation density and $\Omega_f$ the density that drives the current accelerated universe expansion. In the Standard Model the last term reduces to $\Omega_f f(z)=\Lambda$. Clearly the class of cosmological models to test (or the given equations of state) can be reconstructed, in principle, starting from $f(z)$( for details see Sec. 3 of \cite{Benetti:2019gmo}. It is then possible to calculate the predictions of the various cosmological models and to compare them with the data by a Markov chain Monte Carlo process. This approach was applied to a second order Pad\'e polynomials (see Fig. \ref{fig_benetti}) in order to analyse data at both small and large scales \cite{Benetti:2019gmo}i.e. by using recent $H(z)$ measurements from the 1048 measurements, in the redshift range $0.01-2.3$, of the Supernovae Pantheon sample \cite{Scolnic:2017caz}, the galaxy clustering (DES sample) that combine galaxy clustering and weak gravitational lensing measurements, using 1321 square degrees of imaging data \cite{Abbott:2017smn} and the early universe data i.e. the Cosmic Background Radiation anisotropies by Planck Collaboration and the Baryonic Acoustic Oscillation measurements \cite{Aghanim:2019ame,Beutler:2011hx,Ross:2014qpa,Alam:2016hwk}.

A dependence of the $\Omega_m$ value with the cosmographic parameters emerges in the analysis of Ref. \cite{Benetti:2019gmo}, indicating the importance of not fixing this parameter in cosmographic analyses. Furthermore, the authors show an improvement on the parameter constraint using the Cosmic Background Radiation temperature anisotropy power spectrum. 
Moreover, it has been found that the cosmographic series truncated at third order (i.e., one uses just two cosmographic parameters: the deceleration and the jerk), better describe the current observations with respect to the minimal $\Lambda$-CDM cosmological model, showing a better $\Delta\chi^2$ with respect to the minimal standard cosmological model (see \cite{Benetti:2019gmo}).
This result acquires great significance when one recalls that General Relativity is a second order theory. Indeed, the evidence that the third order is required in cosmography to fully describe the data could be interpreted as that a higher order theory is needed to correctly describe the recent expansion of the Universe. In other words, $\Lambda$CDM may be a coarse-grained model that needs higher-order corrections in order to reproduce in a self-consistent way the existent phenomenology.

 Furthermore, it is possible to go beyond General Relativity \cite{cosmo_constant} and the related problems of the (homogeneous and isotropic) $\Lambda$CDM model that includes a cold dark matter (CDM) component and a no-evolving dark energy term with $w=-1$. This model provides the best fit to most of the available cosmological data \cite{Aghanim:2018eyx} at present epoch but it is plagued with conceptual and theoretical shortcomings \cite{Weinberg89}. This state of art may indicate that General Relativity has to be, in some sense, "improved" at cosmological scales \cite{rocco_review}. To this purpose one can look to the \textit{Extended Theories of Gravity} representing a semi-classical approach, in which General Relativity is recovered in the weak energy limit \cite{ETG}. Within this approach, beside the common \textit{metric formalism}, one can also apply the \textit{Palatini formalism} in which the field equations are obtained through the variation of the gravity action with respect to both metric and connection, treated as independent variables. This provides different field equations with respect to those of General Relativity when non-linear terms in the Ricci scalar $R$, or scalar fields non-minimally coupled to gravity are present \cite{Palatini}. One can adopt \emph{a priori} the function of the Ricci scalar $f(R)$ but, in cosmographic context, it is possible to reconstruct the $f(R)$ function without assuming any specific functional form \cite{rocco_pade,rocco_palatini}. The free parameters of the models are therefore obtained by observational data. The results indicate slight departures from the $\Lambda$CDM model, with a time-evolving dark energy term \cite{rocco_chebyshev}.

Therefore, better performing cosmographic series and more precise data at low redshift by future experiments \cite{Levi:2019ggs, Jha:2019rog, Schlegel:2019eqc}, will help toward the final goal to reconstruct the actual cosmic history, to test the effectiveness of General Relativity at any redshift and to investigate the DM and DE mysteries. Indeed, cosmography allows us not only to avoid to choose a-priori a particular cosmological model, but also to reconstruct it by an {\it inverse scattering procedure} by which it is possible to obtain reliable equations of state. 
 
From the current state of the art of this topic it can be concluded that the standard cosmological model shows significant gaps in describing low redshift data, and this is also confirmed using higher-redshift probes such as quasars, as shown in the following session. 
 
\vskip 0.6cm
\section{Constraining dark energy models at high redshifts}
\label{h-z dark energy}

\vskip 1.2cm

 Over the last decades, observations of high-redshift supernovae of type Ia revealed the current accelerated expansion of the Universe (\cite{perl98, per+al99, Riess,Union2.1}). This result has been confirmed by temperature anisotropies of the cosmic microwave background radiation (\cite{PlanckXIII},\cite{Planck15}), and by many different data. According to the most recent estimates, dark energy provides about 75\% of
matter-energy content in the universe. 

The nature of dark energy is unknown. Models of dark energy include, the standard one with 
the cosmological constant having a constant equation of state, $w$, (in this case: $w=-1$), but also one or more of the following: 1) a potential energy of some scalar field \cite{Peebles88a,Peebles88b,Copeland06}, 2) effects due to a possible inhomogeneous distribution of matter or to errors in averaging procedures 3) the standard GR being substituted by extended theories of gravity where dark energy can be described in terms of scalar non minimally coupled scalar fields (\cite{clark},\cite{mdl} and references therein).
Moreover, these scalar fields might be not fundamental and consist of fermion condensates \cite{weinberg}. The effects of these condensates on the expansion of the Universe can be seen in \cite{CCFPV}. There are also other possibilities for dark energy see \cite{esterGL, esterscalar, esterft}, and \cite{sante_interaction2018}. Interacting dark matter/dark energy models have 
been recently introduced in different contexts~\cite{esterint20, amendola, Bonometto17,Bonometto19}. 
In all these cases, in general, $w\not= -1$ and it is not constant with redshift. 

Extracting the information on the effective equation of state $p_\Lambda =w \ \rho_{\Lambda}$ 
of dark energy from observational data is, therefore, at the same time a fundamental problem and a challenging task in Cosmology and Physics: new, independent measurements of the rate of expansion of the Universe, especially in different ranges of redshifts, can provide us with a non-trivial test of the standard cosmological model (see \cite{Riess18a,Riess18b}). 

Other cosmological probes of distances have entered the game. Ghirlanda et al. (2004), Fermiano et al.( 2005) used long GRBs (gamma ray burst) to build up the Hubble diagram up to high redshifts ($z\simeq 9$) (\cite{Ghirlanda04}, and \cite{Firmani05}). GRBs are sudden bursts of hard X–ray / soft gamma–ray emissions, and can be as short as tens of millisecond and as long as several minutes. They are classified as long GRBs if the pulse lasts more than $2$ secs, and show a non-thermal spectrum that generally peaks at few hundreds keV. These objects release enormous amounts of energy (the isotropic equivalent radiated energy is of order of $10^{52}-10^{54}$ erg) and therefore are visible up to very high redshifts being so good candidates as cosmological probes.
Their non-thermal spectra can be modeled with the Band function $N(E)$, a broken power law with the following parameters: the low–energy spectral index $\alpha$, the high energy spectral index $\beta$ and the {\it roll–over} energy $E_0$:

\begin{equation}
N(E)=\left\{
\begin{array}{ll}
 A \left(\frac{E}{100keV}\right)^{\alpha} \exp{\left(\frac{-E}{E_0}\right)}\,,&\left(\alpha-\beta\right)\geq 0\,,\\
 A \left(\frac{\left(\alpha-\beta\right)E}{100keV}\right)^{\alpha-\beta} \exp{\left(\alpha-\beta\right)\left(\frac{E}{100keV}\right)^{\beta}}\,,&\left(\alpha-\beta\right)E_0\leq E\,,\\
\end{array}
\right.
\label{band}
\end{equation}

The spectra have a peak in correspondence of the photon energy $E_{\rm p} = E_0 (2 + \alpha)$. For GRBs with measured spectrum and redshift it is possible to evaluate the intrinsic peak energy, $E_{\rm p,i} = E_{\rm p} (1 + z)$ and the isotropic equivalent radiated energy
 \begin{equation}
E_{\rm iso}= 4 \pi d_L(z,{\mathrm \theta}) \left(1+z\right)^{-1}\int^{10^4/(1+z)}_{1/(1+z)} E N(E)
dE\,.
\label{eiso}
\end{equation}

It turns out that both $E_{\rm p,i}$ and $E_{\rm iso} $ span orders of magnitude and their distributions can be approximated by Gaussians plus a tail at low values of energy. In 2002, for long GRBs the Amati relation was discovered: $E_{\rm p,i}$ was found strongly correlated with $E_{\rm iso}$ (\cite{amati02}): 
\begin{eqnarray}
 \label{eqamatievol2}
&&\log \left[\frac{E_{\rm iso}}{1\;\mathrm{erg}}\right] = b+a \log \left[
 \frac{E_{\mathrm{p,i}}}{300\;\mathrm{keV}} \right]\nonumber\,.
\end{eqnarray}

This correlation can be properly calibrated and allows us to standardize the GRBs as a distance indicators in a way similar to that used for the Phillips relation concerning SNeIas (\cite{Amati08} and reference therein). Then, once the correlation between $E_{\rm p,i}$ and $E_{\rm iso}$ is calibrated, it is possible to estimate the luminosity distance $ d_L(z,{\mathrm \theta})$ by using Eq. (\ref{eiso}), and therefore to build up 
the Gamma Ray Burst Hubble diagram \cite{esterhz14,Megrb11,MGRB2b,MGRB1,MEC11}. 
Furthermore, Eisenstein et al. (2005) introduced BAOs as standard rulers (\cite{Eisenstein05}),
Chavez (2016) used HII galaxies (\cite{Chavez16}), and Negrete et al. (2017) used extreme quasars \cite{Negrete17}.
Recently Risaliti and  Lusso \cite{Lusso16, risaliti1} found a physical relation between the ultraviolet (at $ 2500 \AA$, $L_{UV}$) and the X-ray (at $ 2 \ keV$, $L_{X}$) emissions in quasars. This correlation has an intrinsic dispersion between 0.35 to 0.4 dex. By eliminating quasars with host galaxy contamination, reddening, X-ray obscured objects and radio loudness (\cite{Lusso16}) the dispersion decreases to 0.21-0.24 dex. This correlation provides a way to standardize quasars as distance indicators similar to SNeIa's.

One investigates the dynamical evolution of dark energy by parametrizing the equation of state, $$\displaystyle w(z)=\frac{p(z)}{\rho(z)}\,,$$ assuming some analytical form of $w$ as function of $z$ \cite{esterhz14}. Otherwise, we can follow a cosmographic approach relying upon quantities that are not model dependent. 
In \cite{Lusso16},\cite{risaliti1} a tension appears between the $\Lambda$CDM model and a Hubble diagram extended behind the supernovae of type Ia, built up by using a quasar sample with high-quality ultraviolet and X-ray flux measurements. This sample spans a huge 
range in red-shift, $0.04<z<5.1$. In \cite{Lusso19}.
By adding other 162 Gamma Ray Bursts data, with an even larger range in redshift: 
$0.03<z<9.3$, the existence of such tension has been reinforced (see Fig. (3) in \cite{Lusso19}). The fitting cosmographic models deviate from the standard $\Lambda$CDM model at $>4\sigma$ by using the high-redshift Hubble diagrams of SNe Ia, quasars and GRBs. Since the diagrams of quasars and GRBs are completely independent, the found tension with the standard model is unlikely to arise from unknown systematic effects and seems to imply new physics. Moreover, such tension is confirmed also in \cite{MGRB2b,MGRB1} where the Gamma Ray Burst Hubble diagram was analyzed. 

These results suggest investigating the $\Lambda$CDM tension by assuming several well known cosmological scenarios with different dark energy equation of state and comparing them by means of the Akaike Information Criterion \cite{akaike}. These are: 
\begin{itemize}
\item[(i)] a model with an evolving equation of state empirically parametrized, the CPL model
\cite{cpl1}, \cite{cpl2} see Eq.(\ref{heos}) in Appendix D. 
\item[(ii)]a model where dark energy is due to a self-interacting scalar field, $\varphi$, which drives the acceleration. We consider the potential $V(\varphi) \propto \exp\left\{-\sqrt{\frac{3}{2}}\varphi\right\}$ \cite{MEC11,Megrb11}. For this model exact solutions of the cosmological equations are known, and are shown in the Appendix C: it turns out that all basic cosmological functions surprisingly depend on a single parameter, $\mathcal{H}_0$, which is the Hubble constant when we use the age of the universe as a unit of time.
\item[(iii)] Early dark energy models, where a non-negligible fraction of dark energy exists at early stages of the Universe. 
\end{itemize}
In these models the 
 dark energy density parameter, $\Omega_{DE}$, depends on the present matter
fraction $\Omega_m$ dominated by the dark component, the early dark energy density parameter $\Omega_e$, and the present dark energy equation of state parameter $w_0$ \cite{doran06}. The form of $\Omega_{DE}$ and the Hubble function of these models are provided in Appendix C by Eqs. (\ref{edep}) and (\ref{Hede}).
The redshift investigation uses the Supernovae Cosmology Project Union 2.1 compilation, containing 580 SNIa, spanning the redshift range
$ 0.015 \leq z \leq 1.4.$, a Gamma Ray Bursts
(GRBs) Hubble diagram, consisting of 193 objects, a set of 28 independent measurements of the Hubble parameter, compiled in \cite{farooqb}, and several baryon acoustic oscillations
(BAO) measurements compiled in \cite{Aubourg15}.
The authors obtain the probability distributions of the cosmological parameters for each of the competing models. Their region of confidence is obtained by maximizing the appropriate likelihood functions by using the Monte Carlo Markov chains method. 

\begin{figure}[ht]
\label{FigEster}
\begin{center}
\includegraphics[width=10cm]{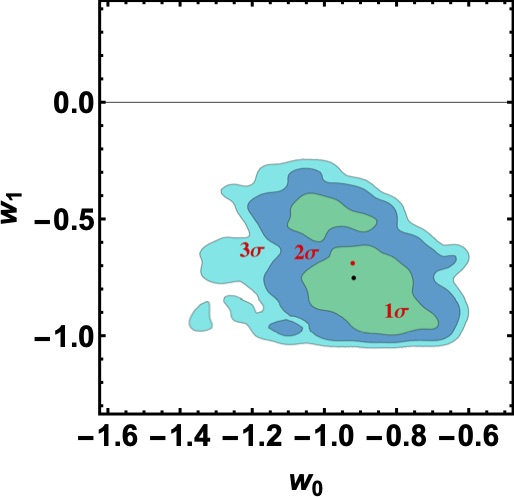}
\end{center}
\caption{2D confidence regions in the $w_0-w_1$ plane for the CPL model, obtained from our full data set. The red points correspond to the best fit value and the mean, respectively.}
\end{figure}

In Tab. (\ref{tab}) of Appendix A are shown the results of our analysis. In Fig. (3) we plot the 2-dim confidence regions in the $w_0-w_1$ plane for the CPL model,obtained from the full data set. The $\Lambda$CDM model corresponds to the case $w_0=-1\,, w_1=0$, and is out at $> 3 \sigma$. 
It is worth noting that this result adds value to that obtained in Sec. 2 showing that third order cosmographic series describe the currently available cosmological observations better than the standard $\Lambda$CDM cosmological model. To bring out more clearly this result one plots, in Fig. (4), the counterpart of the confidence regions in Fig. (3) in terms of cosmographic parameters. We actually implemented a standard cosmographic series in the auxiliary variable $y =\displaystyle \frac{z}{1+z}$, which maps the z-range [ $0,\infty$] into the y-range [$0, 1$]. It turns out that the value $j_0=1.0$ is ruled out at 3 $\sigma$.

We then apply to the above data the Bayesian Akaike criterion and found that among the three models the evolving dark energy, described by the exponential scalar field potential (EXP), is favoured. Let us notice that such evolution of the Dark Energy reverberates also in the Dark Matter component, whose evolution of perturbations with redshift, will now differ from the $\Lambda$CDM case.
\begin{figure}[ht]
\label{FigEster2}
\begin{center}
\includegraphics[width=10cm]{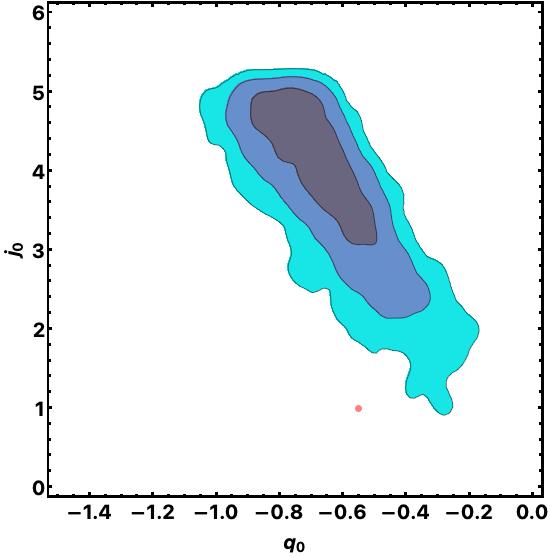}
\end{center}
\caption{2D confidence regions in the $q_0-j_0$ plane obtained from our full data set. For  $\Lambda$ CDM we have: $j_0=1$, and $q_0=-1+\frac{3}{2}\Omega_m$. The pink point corresponds to $\Omega_m=0.3$.}
\end{figure}
Future missions such as the THESEUS observatory for 
Gamma Ray bursts will substantially 
increase the number of Gamma Ray bursts in their Hubble diagram, therefore, providing us with the opportunity to confirm these results. Reliable high redshift cosmological probes alongside with the cosmography approach might indicate the actual dark energy model and the correct gravity theory.
\vskip 1.2cm
\section{The scattering angle of a two-body system}
\label{scatt}
\vskip 0.6cm

 This research issue covers one of the most important topics in classical general relativity today.
In fact, the last few years have witnessed the beginning of the era of gravitational-wave astronomy, with the detection of the 
first signals by the LIGO and Virgo interferometers (see e.g., the web sites 
https://www.ligo.caltech.edu/ and http://www.virgo-gw.eu/).
These detectors have the potential 
to accurately measure the two-body dynamics up to the region where the gravitational
interaction is very strong, so that there is now a pressing need
 to go beyond the post-Newtonian 
approximation by combining results from different approaches.
The Hamiltonian of a two-body system takes information (encoded in several building-block potentials) 
from any theoretical computation of gauge-invariant quantities. 
After having studied several orbital invariants like the redshift and the periastron advance,
all recent literature is mainly focusing on the scattering angle in hyperbolic-like orbits.
This is particularly promising in view of the bridge recently discovered between the classical scattering 
angle in post-Minkowskian context \cite{Damour:2016gwp,Damour:2017zjx} 
and the quantum scattering amplitudes for the same problem, computed by a completely different method of high-energy physics.
The reached accuracy is now limited to the third post-Minkowskian order \cite{Bern:2019nnu,Bern:2019crd} and to the local part of the associated Hamiltonian, but the aim is to go beyond this level, including gravitational self-force information and to study the effects related to the 
non-local (non-trivial) part of the Hamiltonian. 

The topics included in this research involve, besides classical GR, also effective field theories and high-energy 
physics, with a direct superposition with other research interests.
More in detail: in the gravitational interaction of a two-body system, i.e., the capture of one body by the other 
as well as the hyperbolic scattering, two different scenarios must be considered, in view of the 
detection of the associated gravitational wave signals.
While several analytical and semi-analytical methods have been developed so far to study the 
first situation (\lq\lq capture"), this is not the case for the second one (\lq\lq scattering"). 
A renewed interest in the scattering problem has recently emerged with 
studies involving post-Newtonian (PN) and post-Minkowskian (PM) expansions of the scattering angle. The latter has been computed within the post-Newtonian theory at the fourth post-Newtonian level of accuracy in \cite{Bini:2017wfr}, including both local and non-local (tail) contributions, whereas the 
general framework of the post-Minkowskian treatment of the scattering problem has been established in Refs. \cite{Damour:2016gwp,Damour:2017zjx}.
 The level-3 post-Minkowskian Hamiltonian, describing the conservative dynamics of a two-body system, has been 
obtained in \cite{Bern:2019nnu,Bern:2019crd}, where the 
two-loop scattering amplitude of the process and the associated scattering angle at third post-Minkowskian level were computed.
This result in the high-energy limit can be properly compared with previous results \cite{Amati:1990xe}.
In an effort to clarify this situation, Ref. \cite{Bini:2019nra} 
suggested a novel approach, aiming at  improving the current knowledge of the post-Newtonian Hamiltonian of a two-body system. This approach combines various different  theoretical formalisms: post-Newtonian, post-Minkowskian, multipolar-post-Minkowskian, gravitational self-force,
and effective-one-body.
 \cite{Bini:2019nra} has given an independent check of the level-3 post-Minkowskian result, limited at the fifth
post-Newtonian level of accuracy, in terms of the scattering angle $\chi_3$, i.e.,
\begin{eqnarray}
\chi_3&=& -\frac{1}{3 p_\infty^3}+\frac{4}{p_\infty}+(-8\nu+24) p_\infty
+\left(-36\nu+\frac{64}{3}+8\nu^2\right) p_\infty^3
\nonumber\\
&+&\left(-\frac{91}{5}\nu+34\nu^2-8\nu^3\right) p_\infty^5
+\left(\frac{69}{70}\nu+\frac{51}{5}\nu^2-32\nu^3+8\nu^4\right)p_\infty^7 + o(p_\infty^7)\,,
\end{eqnarray}
where the label 3 refers to the third post-Minkowskian level, $\nu=m_1m_2/(m_1+m_2)^2$ denotes the symmetric mass ratio of the two bodies and the relation between the momentum at 
infinity and the (effective) energy of the process is given by 
$p_\infty\equiv \sqrt{{\widehat{\mathcal E}_{\rm eff}}^2-1}$.
In the above expression, the last term proportional to 
$p_\infty^7$ is exactly the level-5 post-Newtonian contribution to $\chi_3$. 
In addition, Bini et al. have anticipated the level-5 post-Newtonian-accurate value for the level-4 post-Minkowskian scattering angle 
$\chi_4= \chi_4^{\rm loc} + \chi_4^{\rm nonloc}$, with the local contribution 
$\chi_4^{\rm loc}(p_\infty)$ given in Appendix \ref{App:scatt}, Eq. (\ref{eq:chi4_loc}). 

Finally, the analysis developed in \cite{Bini:2019nra} has allowed to identify the local part of the level-5 post-Newtonian Hamiltonian modulo two 
unknown functions, $a_6^{\nu^2}$ and $\bar d_5^{\nu^2}$ that, however, are of the second order in the 
symmetric mass-ratio of the two bodies, see Appendix \ref{App:scatt}.

Further studies and possible generalizations are currently being considered \cite{Damour2019,BDG_2019,Kalin:2019inp,Kalin:2020mvi,Bern:2020buy}, including both the 6PN local and non-local contribution to the scattering angle.

We point out that measuring the scattering angle in a two-body system implies the determination of (gauge-invariant) information about the radiated gravitational wave energy along the full scattering process. Therefore, the analytical modeling for this quantity is important in the analysis of future detections of gravitational wave signals by gravitational wave interferometers. Noticeably, an accurate investigation of the merging of two BHs may yield to a precious test about them being the DM in galaxies.

\vskip 1.2cm

\section{Classical physics in curved spacetime backgrounds}

\vskip 0.6cm

 {\it (i) Scalar wave equation and Ermakov-Pinney.} 
 
The field equations of classical field theory in curved spacetime, i.e., scalar wave equation, Maxwell equations and Einstein equations, are wave equations with variable coefficients, whose solution is known only in a few cases. Thus, a systematic technique for dealing with such hyperbolic equations may lead to a valuable physical insight.

In the work in \cite{EM2019},
the first part of the paper proves that a subset of the Ermakov-Pinney 
equations can be obtained by differentiation of a first-order non-linear differential
equation. The second part of the paper proves that the equation for the
amplitude function for the parametrix of the scalar wave equation can be obtained by
differentiating covariantly a first-order non-linear equation. The construction of
such a first-order non-linear equation relies upon a pair of auxiliary $1$-forms
$(\psi,\rho)$. The $1$-form $\psi$ satisfies the vanishing divergence condition
${\rm div}(\psi)=0$, whereas the $1$-form $\rho$ fulfills the non-linear equation 
$${\rm div}(\rho)+\langle \rho, \rho \rangle=0.$$
The auxiliary $1$-forms $(\psi,\rho)$ are evaluated
explicitly in Kasner spacetime. Therefore, amplitude and phase function in the parametrix are obtained. The novel method developed in this paper can be used in studying the electromagnetic field of binary systems in relativistic astrophysics. 

The key property to be used is the fact that the technical 
difficulty of dealing with a coupled set of partial differential equations for the
electromagnetic potential can be overcome by exploiting the Hertz potentials, that lead 
eventually to a linear wave equation for a complex scalar field
\cite{CK1974}. The real and imaginary
part of such a scalar field are then subject to a linear wave equation, and the original method relying upon two auxiliary $1$-forms may come into play. 

In \cite{BE2019}, the authors have solved the 
Ermakov-Pinney equation associated to the scalar wave equation in Schwarzschild, de Sitter
and gravitational-wave spacetimes. For example, in the Schwarzschild case it is found
that a positive coupling constant leads to spatially damped oscillations of the field (see Fig. 5), whereas a
negative or vanishing one is associated with a blow-up of the solutions. 
\begin{figure}
 \centering
 \includegraphics[scale=0.73]{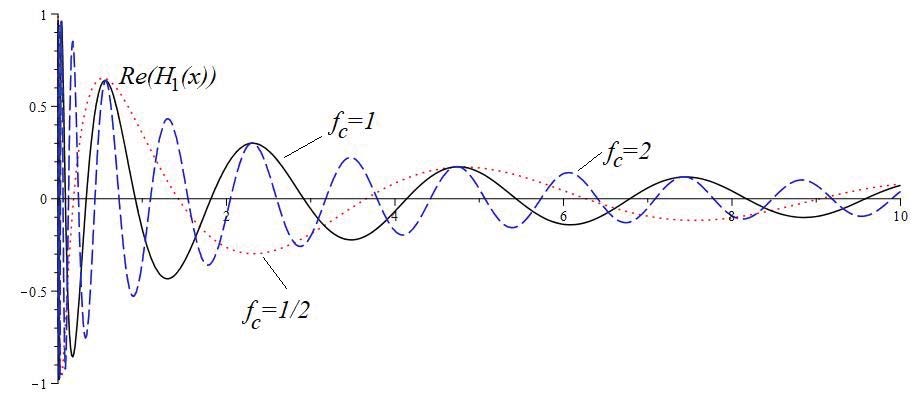}
 \caption{Amplitude function for a timelike current in Schwarzschild space-time.  The real part of the Heun function $H_{1}(x)$ is plotted, which contributes to the amplitude for different values of the coupling constant $f_{c}=\frac{1}{2},1,2$. All curves show a spatially damped oscillating behaviour with a wavelength qualitatively proportional to the inverse of the coupling constant. The imaginary parts of the amplitude have a similar behaviour.}
 \label{Fig_esposito}
\end{figure}
In light of this, and bearing in mind the interest of studying in greater detail the quantization of
fields, it seems more important to focus on the case of a positive coupling constant, a 
topic to be developed in future work. It would also be valuable to understand the
relation (if it exists) with the asymptotic behaviour of solutions of the scalar wave equation 
discovered by Schmidt and Stewart \cite{SS1973}.
\vskip 0.2cm
\noindent

{\it (ii) Trapped surfaces in general relativity}. 
After the proof by Christodoulou and Klainerman of the global non-linear stability of
Minkowski spacetime \cite{CK1993}, Christodoulou decided to apply his optical formalism in order to
prove that, in the non-linear equation that describes the formation of trapped surfaces, some terms can be omitted while keeping under control the resulting error, that remains suitably small \cite{C2009}. More recently, in order to deal with exact
equations, without any approximation, in 
 \cite{BE2018} trapped surfaces have been studied instead
from the point of view of local isometric embedding
into three-dimensional Riemannian manifolds. When a two-surface is embedded into three-dimensional
Euclidean space, the problem of finding all surfaces applicable upon it gives rise to a
non-linear partial differential equation of the Monge-Amp\`ere type, first discovered by Darboux \cite{DA}, and later reformulated by Weingarten \cite{WE}. Even today, this problem remains very difficult, despite some remarkable results. In \cite{BE2018}, the authors have found an original way of
generalizing the Darboux technique, which leads to a coupled set of six non-linear partial 
differential equations. For the $3$-manifolds occurring in 
Friedmann-(Lemaitre)-Robertson-Walker cosmologies, they have shown that the local isometric embedding
of trapped surfaces into such cosmologies can be proved by solving just one non-linear equation.
Such an equation has been solved for the Friedmann models associated with positive, zero, negative curvature of the spatial sections, respectively.
However, the task of solving such a non-linear equation in the most general case remains extremely difficult, and deserves a substantial effort, because there is a proliferation of terms in the non-linear equation, without any simplifying ansatz.

\vskip 1.2cm

\section {GR as the bridge between classical and quantum worlds}

 \vskip 0.6cm
 
 Motivated by recent efforts to apply the asymptotic symmetry group of asymptotically flat space-times to the investigation
of black holes' soft hair \cite{HPS2016}, 
in \cite{EA2018} the authors remark that
half of the Bondi-Metzner-Sachs (BMS) transformations are made of orientation-preserving homeomorphisms of the extended complex plane known as fractional linear (or M\"{o}bius) transformations. These can be of four kinds, i.e., they are
parabolic, or hyperbolic, or elliptic, or loxodromic, depending on the
number of fixed points and on the value of the trace of the associated $2 \times 2$ matrix
in the projective version of the $SL(2,{\bf C})$ group. The resulting particular forms of
$SL(2,{\bf C})$ matrices affect also the other half of BMS transformations. They 
have been used in \cite{EA2018} to propose four 
realizations of the asymptotic symmetry group that are called, again, parabolic, or hyperbolic, or elliptic, or loxodromic. 

 Moreover, it has been proved that a particular subset of hyperbolic and loxodromic transformations, i.e., those that have a trace that approaches $\infty$, correspond to the fulfillment 
of limit-point condition for singular Sturm-Liouville problems. 
Thus, a very deep link might exist between the language for describing asymptotically flat spacetimes
and the world of complex analysis and self-adjoint problems in ordinary quantum mechanics. In other words, we are suggesting that
General Relativity plays the role of gateway that establishes a correspondence between classical and quantum world. Such a viewpoint
is of particular interest for the theoretical analysis of fundamental interactions.
 
So far, relativists have never considered the possibility of defining the concept of boundary in a measure-theoretic way.
On the other hand, in a branch of mathematics known as geometric
measure theory, the usefulness has been discovered long ago of yet another concept, i.e. the reduced
boundary of a finite-perimeter set. In \cite{EMe2019} 
the authors have proposed therefore a definition 
of finite-perimeter sets and their reduced boundary in general relativity. Moreover, an integral formula of geometric measure theory has been evaluated explicitly in the relevant case of Euclidean Schwarzschild geometry, for the first time in the literature. This prepares the ground for a measure-theoretic approach to several concepts in gravitational physics, supplemented by geometric insight.

Interestingly, some observational tests of quantum gravity in the weak-gravity
regime has become, in principle, conceivable by applying the effective-gravity picture
of Donoghue \cite{D1994}. In particular, within the
framework of the solar system, approximate realizations of the three-body problem occur when a comet
approaches a planet; such a configuration
was investigated by Tisserand \cite{TI} in Newtonian gravity. The exact relativistic
treatment of the problem is not easy, but the work in \cite{BET2019} develops an approximate computational scheme which evaluates for the first time the very tiny effective-gravity
correction to the equation of the surface for all points of which it is equally legitimate to regard
the heliocentric motion as being perturbed by the attraction of Jupiter, or the jovicentric
motion as being perturbed by the gravitational attraction of the Sun.
In the second part of \cite{BET2019}, comet trajectories for effective and Newtonian gravity are obtained and compared in detail. 
In the years to come, the relativistic analysis of cometary motions will require the inclusion of high orders of the post-Newtonian formalism, a task extremely difficult for three-body systems. This investigation will lead to a better understanding of the occurrence of chaotic behaviour in general relativity. 

Remarkably, such behaviour is very distinct from the eventual effects, in the Solar System, of alternatives to the DM that, on the other hand, can affect the cometary motions at the level of the above high orders.

\vskip 1.2cm

\section{Quantum Field Theory in curved space-time and neutrino physics}

\vskip 0.6cm
\label{Luciano}
The search for a merging between General Relativity and Quantum Field Theory is the most demanding task of theoretical physics nowadays. Notwithstanding the zoo of models developed in recent years,
the problem of finding a consistent theory of quantum gravity is still open, due to the lack of experimental guidance. 
Groping in the dark, significant progress is then expected to be achieved via indirect investigation. For instance, in \cite{Mavr} it was argued that quantum gravity may involve foam models endowed with stochastic
fluctuations of the space-time background. One of the most sensitive probes of these fluctuations 
could be \emph{neutrinos} 
through decoherence effects on \emph{flavour oscillations}. 
Similarly, in \cite{VedralMarletto} neutrinos were selected as best candidates to
witness the gravity-induced entanglement between massive systems, being only affected by the weak and gravitational interactions.

On the other hand, in the cosmological framework it has been widely discussed \cite{SFU,SFU2,SFU3} the active r\^ole played by primordial neutrinos
in the formation of large-scale structures (LSS) of the Universe and, in particular, in giving a power at a large scale that otherwise could not be explained in the ordinary
cold dark matter scenario at the critical density. Due to 
the considerable impact of neutrinos on the process of structure formation, observations of matter clustering in different epochs of the Universe have then contributed to build a solid bridge between cosmology and neutrino physics. This has also enabled us to put very competitive constraints on characteristic neutrino parameters, such as the neutrino mass sum $\sum_i m_{\nu_i}<0.72\,eV$ at 95\% CL (through measurements of large scale polarization and CMB temperature anisotropies) or $\sum_i m_{\nu_i}<0.14\,eV$ at 95\% CL (including measurements of Lyman-$\alpha$ forests), and 
the number of relativistic species at recombination $N_{eff}=3.03\pm 0.18$ at 68\% CL (exploiting LSS and CMB data). More details can be found in Ref. \cite{Lattanzi} and therein.

In light of the above discussion, it is believed that a deeper study of the fundamental nature of neutrinos could not only improve our knowledge of particle physics beyond the Standard Model, but also provide novel insights into a series of related problems, the most prominent ones being the development of a quantum theory of gravity and the understanding 
of the formation of the current Universe from small early density fluctuations.

Neutrinos are the most puzzling particles currently known.
Although a series of findings have confirmed Pontecorvo's idea of flavour mixing and oscillations, several questions remain unsolved.
Among these, the problem of the very nature of asymptotic states -- flavour or mass -- is still a vibrant subject of analysis \cite{Blas95, Giuntibook, Casimirmix}. Recently,
a challenging test bench has been provided in \cite{Ahluwalia:2015kxa,Blasone:2018czm} within the framework of the weak decay of uniformly accelerated protons (inverse $\beta$-decay). The idea that non-inertial protons
are allowed to decay if exposed to sufficiently large accelerations traces back to \cite{Ginz}. In \cite{Matsas:1999jx} the inverse $\beta$-decay was exhibited as a theoretical proof of Unruh effect \cite{Unruh} for the consistency of QFT in curved space-time. However, in these works neutrinos were simplistically considered as massless. 
The first successful attempt to embed mixing was made in \cite{Blasone:2018czm}, concluding 
that the use of the flavour basis is mandatory both to preserve the general covariance of QFT and to account for neutrino oscillations. 

Starting from the outlined picture, let us review the r\^ole of the superposition of neutrino mass states in the inverse $\beta$-decay.
By employing the $S$-matrix formalism, the scalar decay rate of accelerated protons is calculated in both the inertial and comoving frames in a two-flavour model (the three-flavour description including CP violation effects is contained in \cite{BLLP2020}). Perspectives and possible experimental implications are then discussed. 
 
 %%%%%%%%%%%%%%%%%%%%%%%%%%%%%%%%%%%%%%%%%%%%%%%%%%%%%%%%%%%%%
\begin{figure}[t]
\begin{center}
\includegraphics[angle=0,width=.28\textwidth]{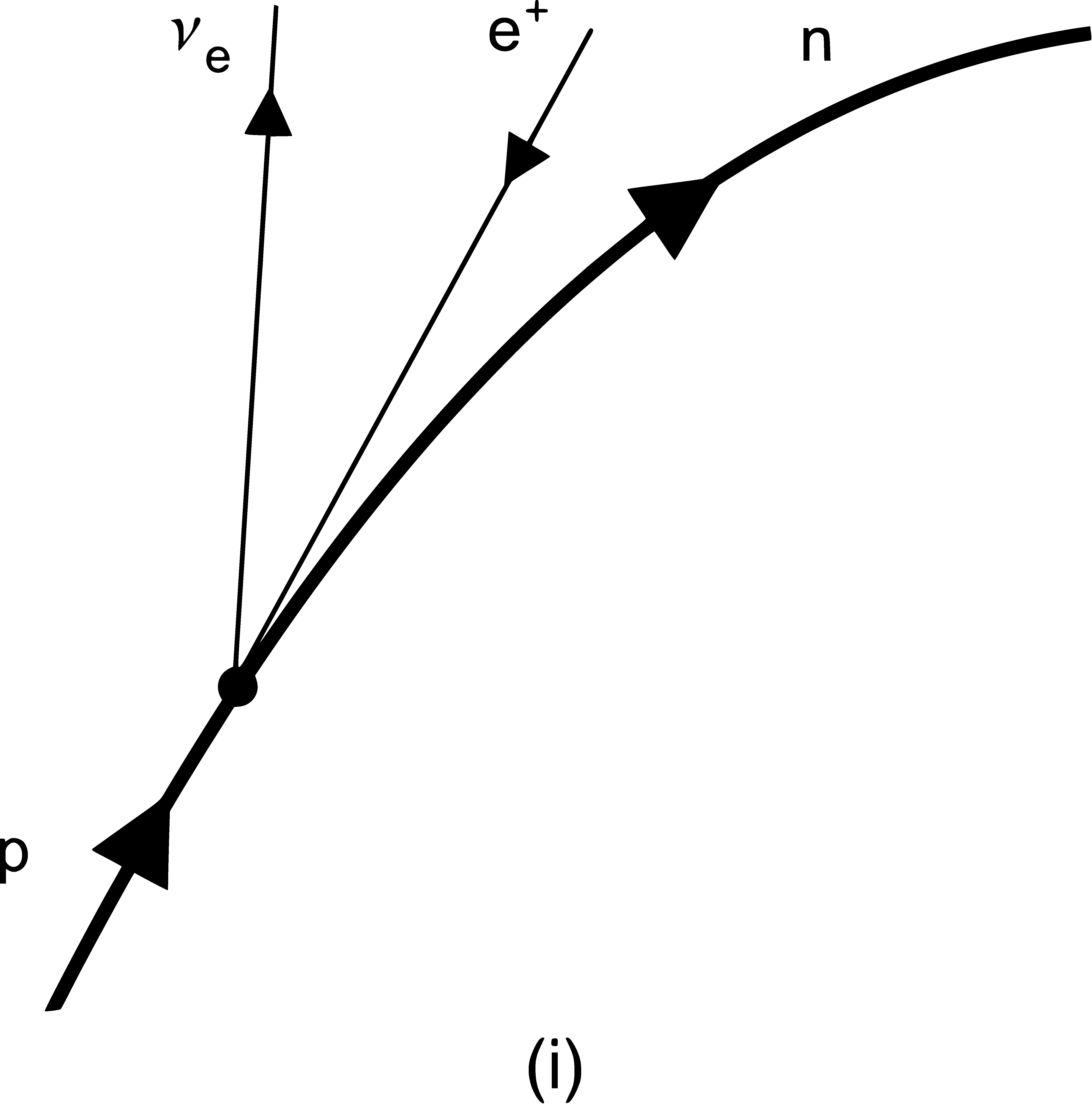}%
\includegraphics[angle=0,width=.71\textwidth]{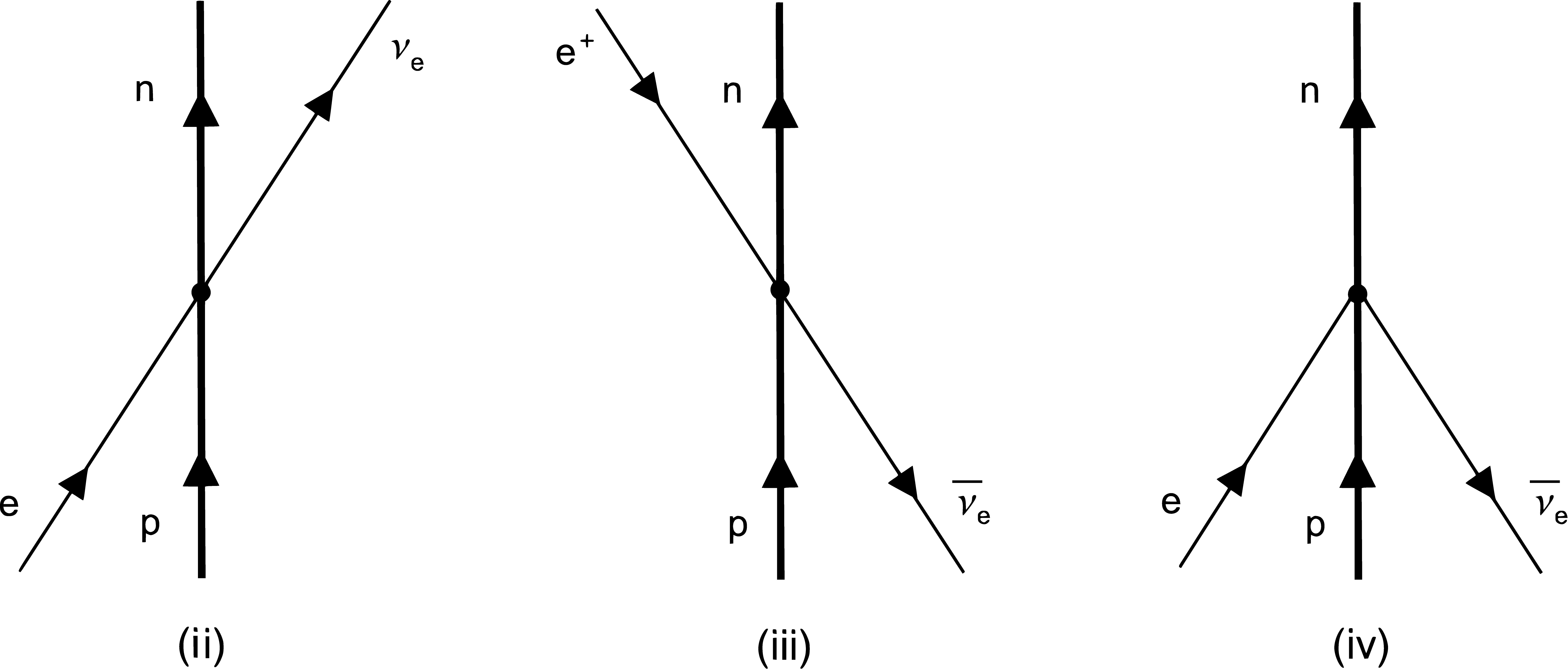}%
\caption{Proton decay in the laboratory (i) and comoving (ii, iii, iv) frames (time is on the vertical axis).}
\label{fig_luciano}
\end{center}
\end{figure}
%%%%%%%%%%%%%%%%%%%%%%%%%%%%%%%%%%%%%%%%%%%%%%%%%%%%%%%%%%%%%%%%%%%
 
\emph{Laboratory frame}. Concerning an inertial observer, the accelerating source supplies the proton with the missing energy to decay as 
$(i)\,\,p\rightarrow n\hspace{0.2mm}+\hspace{0.2mm}
e^{+}\hspace{0.2mm}+\hspace{0.2mm}\nu_e$ (see the diagram $(i)$ in Fig.~\ref{fig_luciano}).
If one assumes 
the acceleration to be much smaller than the masses
of the ${Z^0}$ and $W^{\pm}$,
the weak coupling can be described by a Fermi-like action $S_I$ with a classical (quantum) hadronic (lepton) current \cite{Blasone:2018czm,Matsas:1999jx}. Accordingly, the tree-level transition
amplitude reads $\mathcal{A}^{\mathrm{lab}}\equiv\langle n|\otimes\langle 
e^{+},\nu_{e}|{S}_{I}|0\rangle\otimes|p\rangle$. This leads to the following decay rate
\begin{equation}
\label{lab}
\Gamma^{(\mathrm{i})}\,\equiv\,\frac{1}{T}\int d^3k_\nu\,d^3k_e\sum_{\sigma_e,\sigma_{\nu}}
\big|\mathcal{A}^{{\mathrm{lab}}}\big|^2\,=\,\cos^4\theta\, \Gamma_{1}\,+\,
\sin^4\theta\,\Gamma_{2}\,+\,\cos^2\theta\sin^2\theta\,\Gamma_{12}\,,
\end{equation}
where the Pontecorvo state
$|\nu_e\rangle=\cos\theta|\nu_1\rangle+\sin\theta|\nu_2\rangle$ has been used for the asymptotic neutrino, $\theta$ is the mixing angle and $|\nu_i\rangle$ ($i=1,2$) denote the mass states (the expressions of
$\Gamma_i$ and $\Gamma_{12}$ are given in \cite{Blasone:2018czm}).

Remarkably, in \cite{PLB}
it has been proved that the flavour-violating process
$p\rightarrow n\hspace{0.2mm}+\hspace{0.2mm}
e^{+}\hspace{0.2mm}+\hspace{0.2mm}\nu_\mu$ has a non-vanishing probability as well, due to the
asymptotic occurrence of neutrino oscillations. Accounting for such a behavior further strengthens the use of flavour (rather than mass) states for neutrinos \cite{PLB,Remarks}.

\emph{Comoving frame}. The question now arises as how to describe the above process for an observer comoving with the proton. Although one may envisage that the decay is kinematically forbidden in this frame, this is not the case, since the proton interacts with electrons and anti-neutrinos popping out from the thermal vacuum due to Unruh effect (see the last three diagrams in Fig. \ref{fig_luciano}). Details on the calculation of the decay rate are carried out in Appendix~\ref{INVB}, showing the result to be consistent with Eq. \eqref{lab}. 
Therefore, unlike previous claims in literature \cite{Ahluwalia:2015kxa},
flavour mixing is perfectly consistent with the covariance of QFT, provided that asymptotic neutrinos are assumed to be \emph{flavour states}.
 
 The r\^ole of the non-thermality of Unruh spectrum for mixed fields \cite{Luciano2017,Luciano2017bis,LucJiz} and the effects induced by gravity~\cite{ETG,Ahluwalia:1996ev,Cardall,CQGLuciano,CQGLucianobis} and acceleration~\cite{Capozziello99,Dvornikov15, Blasonepl} on the standard oscillation formula have still to be fully addressed. Worthy of attention are also the entanglement properties that emerge among neutrinos and other decay products in interaction processes. Beyond its inherent theoretical relevance, the tension between flavour and mass states may be of interest at the phenomenological level, as well. Recently, indeed, it has been shown that the predicted spectra of the neutrino capture on tritium and of the tritium $\beta$-decay are sensitive to whether neutrinos interact as massive or flavour eigenstates. Thus, it is expected that the output of experiments such as KATRIN \cite{Katrin} (that aims at measuring the mass of the electron antineutrino by investigating the spectrum of electrons emitted by the $\beta$-decay of tritium) or PTOLEMY \cite{Ptolemy,Ptolemybis} (which is projected to reveal the Cosmic Neutrino Background via capture on tritium), 
may provide important pieces of information in the considered problem, deepening our understanding of neutrino physics. In addition, let us notice that neutrinos might add up to the total dark matter amount of the Universe. In this context, KATRIN and PTOLEMY setups could detect these particles in the keV mass range, that is the scale at which they should form the DM of the Universe. 
Potential connections between neutrinos (and, more general, particle) physics and dark matter are also under the active
investigation of upcoming experiments such as AEDGE~\cite{AEDGE}.

\vskip 1.2cm

\section{Dirac and Majorana neutrinos}

\vskip 0.6cm

 \label{Quaranta}
 
 The fundamental nature of neutrinos has eluded physicists so far. Its determination is important due to the prominent role that neutrinos play in Cosmology. In particular, both the known active neutrinos and the hypothetical sterile neutrinos have been proposed as dark matter candidates, respectively as hot dark matter and warm dark matter \cite{Boyarski2019}.
 The issue suffers from 
a substantial lack of experimental tests, the only attempts are based on the phenomenon 
of neutrinoless double beta decay \cite{Cardani2019}. While Dirac neutrinos have a distinct anti-particle (anti-neutrinos), Majorana neutrinos coincide with their own anti-particle. 

 The most promising way to reveal the possible Majorana nature of neutrinos is the experimental observation
of the double beta decay without neutrinos \cite{ref1} (see Refs. \cite{ref2} for alternative possibilities). 
The main difficulty in determining the Majorana character is that in theories with $V-A$ interactions, this is the case of the Standard Model, 
any observable difference between Dirac and Majorana neutrinos is suppressed by a factor $(m_\mu/E)^2$, being $E$ the energy scale of the process \cite{Ref27}. 
However, an interesting possibility arises in the case in which neutrinos have new interactions beyond the SM \cite{Ref26}. 
In is worth of mentioning, for example, the proposal of Ref. \cite{Ref6} in which the author considers the most general Lorentz-invariant form of neutrino-fermion 
interactions. This includes scalar, pseudo-scalar, vector, axial-vector and tensor couplings. In such a circumstance, it follows that
in elastic neutrino-electron scattering, the ratio of forward to backward scattering cross sections $R_p$ could be used for distinguishing 
Dirac and Majorana neutrinos. More specifically, $R_p\leqslant 2$ for Majorana neutrinos, and $R_p>4$ for Dirac neutrinos.
Other alternative methods to determine the Majorana and Dirac nature stem from the differences that arise in presence of decoherence \cite{Capolupo2019,CapolupoB2020} and in the propagation through a medium~\cite{Capolupo2018}. Let us recall the main distinctions between the two kinds of neutrinos. Let $\nu$ denote the neutrino field and let $\nu_{R,L}= \frac{1 \pm \gamma_5}{2} \nu$ be its left--handed and right--handed components. The Dirac mass term in the neutrino Lagrangian has the form 
\begin{equation}
L_D = \frac{m_{LR}}{2}(\bar{\nu}_L \nu_R + \bar{\nu}_R \nu_L) \, 
\end{equation} with $\bar{\nu} = \nu^{\dagger} \gamma^0$. On the other hand the Majorana mass term has the structure 
\begin{equation}
 L_{M,L} = \frac{m_{LL}}{2} (\overline{(\nu_L)^c} \nu_L + \bar{\nu}_L (\nu_L)^c) \,
\end{equation}
 where the superscript $c$ denotes the charge conjugated field. If right--handed neutrinos are allowed for, one has an analogous Majorana mass in terms of the right--handed components, namely
 \begin{equation}
L_{M,R} = \frac{m_{RR}}{2} (\overline{(\nu_R)^c} \nu_R + \bar{\nu}_R (\nu_R)^c) \ . 
 \end{equation}
 A heavy right--handed Majorana mass $m_{RR}$ might explain the smallness of the mass of the observed neutrinos via the see--saw mechanism (see ref. \cite{King2004} for a review on several neutrino mass models). The different structure in terms of left--handed and right--handed components implies different transformation properties for the two kinds of neutrinos. Indeed, while the Dirac 
Lagrangian is invariant under global $U(1)$ transformations, the Majorana Lagrangian 
is not. As a consequence, Dirac neutrinos preserve the total lepton number, whereas Majorana 
neutrinos allow for lepton number violating processes. In addition, a different number 
of phases appears in the mixing matrix. For $n$ flavours, there are $N_D = \frac{(n-1)(n-2)}{2}$ 
Dirac phases and $N_M = \frac{n(n-1)}{2}$ Majorana phases. This work is focused on $2$-flavour mixing, 
which features a single Majorana phase $\phi$. 
The Majorana mixing matrix admits various parametrizations
\begin{equation}
U_1 = \left(
 {\begin{array}{cc}
 \cos \theta & \sin \theta\, e^{i \phi}\\
 - \sin \theta & \cos \theta\, e^{i \phi}\\
 \end{array}}
 \right) \ \ \ \ \ \ U_2 = \left(
 {\begin{array}{cc}
 \cos \theta & \sin \theta\, e^{-i \phi}\\
 - \sin \theta\, e^{i \phi} & \cos \theta \\
 \end{array}} \right)
\end{equation}
which lead to the same oscillation formulae in absence of decoherence \cite{Giunti,CapolupoC2020,CapolupoD2020}.

The Dirac mixing matrix is in any case obtained for $\phi = 0$. To study propagation with decoherence, it is necessary to 
model neutrinos as open quantum systems, described by means of density matrices $\rho(t)$. The time evolution of $\rho(t)$ is driven by a Lindblad-Kossakowski master equation \cite{Lindblad, Lindblad2}: 
$\frac{\partial \rho(t)}{\partial t} = -\frac{i}{\hbar}[H_{eff},\rho(t)] + D[\rho(t)]$.
The effect of decoherence is contained in the \textit{dissipator} $ D[\rho(t)]$, which for the case at 
hand is a $3 \times 3$ matrix on the space of bounded operators $B(\mathcal{H})$ on the two-level Hilbert space $\mathcal{H}$. 

The elements of the dissipation matrix $D$ are phenomenological 
parameters, to be extracted from the experimental setup. If $ D$ is diagonal, Dirac and Majorana 
neutrinos happen to obey the same oscillation formulae. For a non-diagonal dissipator, 
however, the two are distinct. One considers a dissipator with two off-diagonal components and plugs it in the master equation. Projecting on the $SU(2)$ basis, one obtains a 
linear system for the components $\rho_{\mu} (t) = \mathrm{Tr}(\rho (t) \sigma_{\mu})$ which is easily solved. 
Then the transition probability from flavour 
$\sigma$ to flavour $\varrho$ can be computed as $P_{\nu_{\sigma} \rightarrow \nu_{\varrho}}(t) 
= \mathrm{Tr} \left[\rho_{\varrho}(t) \rho_{\sigma}(0) \right]$. Notice that, at this stage, 
the choice of a parametrization for the mixing matrix becomes relevant, as $U_1$ and $U_2$ 
produce different formulae. In particular, choosing $U_2$ results in probabilities that have an explicit 
dependence on the Majorana phase $\phi$, which prompts an asymmetry between the particle 
and anti-particle transitions. One can introduce the $CP$-violating quantity 
$\Delta_{CP}^{\sigma \rightarrow \varrho}(t) = P_{\nu_{\sigma} \rightarrow \nu_{\varrho}}(t) - P_{\overline{\nu}_{\sigma} \rightarrow \overline{\nu}_{\varrho}}(t)$ as a measure of the $CP$ asymmetry. For Majorana neutrinos, even the survival probability $P_{\nu_{e} 
\rightarrow \nu_{e}}(t)$ shows a $CP$ asymmetry. It should be remarked that this $CP$ violation, 
depending exclusively on the Majorana phase $\phi$, is absent in the standard 3-flavour 
mixing of Dirac neutrinos, and is therefore distinct from the $CP$ violation arising from the Dirac phase $\delta_{CP}$. On the other hand, $\Delta_{CP} = 0$ for any transition 
involving Dirac neutrinos. This can be generalized to 
three flavours in a straightforward way. 
%%%%%%%%%%%%%%%%%%%%%%%%%%%%%%%%%%%%%%%%%%%%%%%%%%%%%%%%%%%%%
\begin{figure}[!]
\begin{center}
\includegraphics[width=0.59\textwidth]{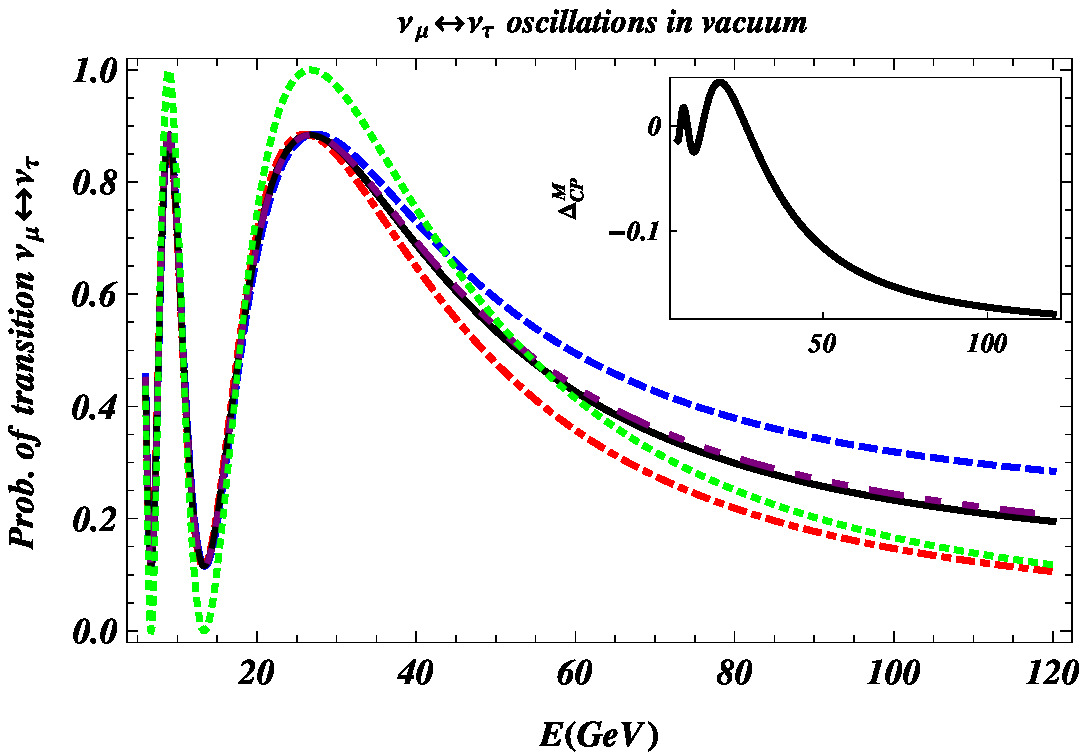}%
\vskip 0.6cm
\includegraphics[width=0.70\textwidth]{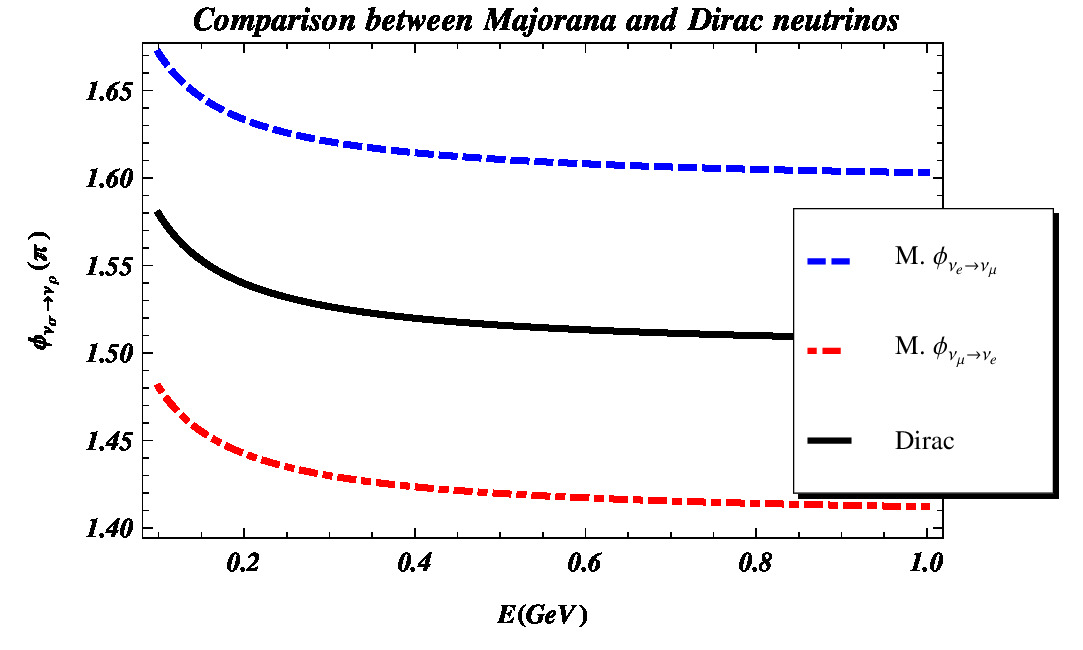}%
\caption{Upper panel: plot of the oscillation formula $P_{\nu_{\mu}
\rightarrow \nu_{\tau}} $ for Majorana neutrinos and for Dirac neutrinos
($\phi = 0$, the black lines), as a function of the energy $E$. The
experimental bounds from long baseline experiments on the decoherence
parameters are used; Lower panel: plot of the geometric phases
$\Phi^{g}_{\nu_{e}\rightarrow \nu_{\mu}} $ (the blue dashed line) and
$\Phi^{g}_{\nu_{\mu}\rightarrow \nu_{e}} $ (the red dot--dashed line) for
Majorana neutrinos as a function of the neutrino energy $E$, for a
distance length $z = 300 km$. The geometric phase
$\Phi^{g}_{\nu_{e}\rightarrow \nu_{\mu}} =\Phi^{g}_{\nu_{\mu}\rightarrow
\nu_{e}} $ for Dirac neutrinos is represented by the black solid line.}
\label{fig_quaranta}
\end{center}
\end{figure}
%%%%%%%%%%%%%%%%%%%%%%%%%%%%%%%%%%%%%%%%%%%%%%%%%%%%%%%%%%%%%%%%%%%
Another distinction between Dirac and Majorana neutrinos comes up from the analysis of geometric phases for neutrinos propagating in matter (see appendix). Using the Mukunda-Simon definition, one 
concludes that the geometric phase associated to a single flavour Eq. \eqref{SameFlavorPhase} is not affected by the 
Dirac/Majorana distinction \cite{Capolupo2018}. On the contrary, the phases associated 
with the mixing, Eqs. \eqref{DiffFlavorPhase1} and \eqref{DiffFlavorPhase2} show an explicit dependence on the 
Majorana phase $\phi$. As for decoherence, we find an asymmetry in the oscillations 
of Majorana neutrinos, which is absent in the Dirac case. The asymmetries in the transition probabilities and geometric phases are exemplified in the figure (\ref{fig_quaranta}). In the upper panel we show the transition probability $P_{\nu_{\mu} \rightarrow \nu_{\tau}} (z)$ for Dirac and Majorana neutrinos when decoherence is taken into account, for a distance $z = 300 km$ as a function of energy. The dissipator parameters are chosen in correspondence with the bounds from long baseline experiments (see \cite{Capolupo2019} and the references therein). In the lower panel we plot the geometric phases associated to neutrino oscillations (see Appendix F) as a function of energy. 

Future studies \cite{Capolupo2020} shall be aimed at finding additional tools and phenomena that can highlight the fundamental nature of neutrinos. The determination of the latter is important to shed light on one of the most elusive particles in the universe, which perhaps constitutes the strongest evidence for physics beyond the standard model of particles and may direct us towards the searched new "dark sector". 

\vskip 1.2cm
 
\section{Quantum fields in curved space-times}

\vskip 0.6cm

\label{Sorge}
 
Several unresolved questions concerning the origin of our Universe, its very evolution and fate, are still waiting for quantum gravity - a theory allowing to manage and reconcile general relativity with quantum mechanics. In that respect, also the apparent presence in the Universe of "dark substances" such as Dark Energy and Dark Matter could play a crucial role. 
One of the difficulties of developing a quantum gravity theory is that quantum gravitational effects are expected near the Planck scale, around $10^{-35}$ meters, so by far out of reach, given the present technology. On the other hand, the approach to quantum cosmology through a suitable quantum-mechanical translation of some known solutions in classical general relativity (and its possible extensions) could perhaps offer a back-door entry towards quantum gravity. Quantum effects involved in the early stages of a cosmological evolution may be of some relevance also at a low-energy microscopical level, influencing the microphysics in a classical background. In that respect,
the study of the behaviour of quantum fields in a curved space-time background represents an 
inescapable first step towards the quest for a comprehensive, self-consistent quantum theory of 
gravity \cite{birrell}. In the semi-classical approach, the background - although allowed to 
evolve in time - is assumed not to be affected by the quantum field itself (i.e., 
back-reaction is not taken into account), the former being only sourced by some given 
{\em classical} distribution of mass-energy and completely described by the Einstein field 
equations. Nevertheless, even in such semi-classical scenario, several non-trivial 
manifestations of the background influence upon a quantum field appear, often related to the 
lack of a privileged reference frame in presence of a curved background. Hence, special attention 
has to be payed to the role of the physical observer's frame, in which the relevant 
measurements are performed. 

Vacuum fluctuations are a distinctive feature pertaining to the quantum nature of a matter field.
One of their most famous manifestations is the well-known Casimir effect \cite{casimir1,lamoreaux,carugno,milton0,bordag2}. Modification in the 
vacuum energy can be also induced by external fields, including gravity \cite{calloni1,fulling1,esposito,nazari}. In such latter case, 
the space-time background and the quantum field confinement may conspire together, giving rise 
to several interesting effects \cite{garufi,avino}. 

Looking at those effects, the Casimir energy has been extensively investigated, taking into 
account a small Casimir apparatus both in the weak and in the strong gravitational field limit. 
Analysis of a Casimir cavity at rest in a weak, static gravitational field has shown 
a small change in the value of the Casimir energy. Such a result has been 
first obtained by using a field mode decomposition technique \cite{Sorge:2005ed}. Recently, the same result has been confirmed by means of a quite different approach based upon the Schwinger 
effective-action method \cite{Sorge:2019ldb}. Furthermore, a similar behaviour has been obtained by considering a 
cavity in free fall into a black hole \cite{Sorge:2018lhw, Sorge:2019ecb, Sorge:ref_2}. The Casimir effect has been investigated also in presence of (weak) gravitomagnetism. 
An interesting result is that - up to the lowest order of approximation - a gravitomagnetic 
field causes no distortion in the vacuum energy \cite{Sorge:2009zz} (the extension to modified theories of gravity has been studied in \cite{caslam1,caslam2,caslam3}). However, a deeper and more 
exhaustive analysis in the field of a Kerr black hole, has revealed that the Casimir energy 
density is indeed sensitive to the Kerr gravitational dragging \cite{Sorge:2014uma}. Actually, a small Casimir apparatus orbiting in the Kerr equatorial plane suffers a change in the Casimir energy 
density. Such a change vanishes in the case of a zero-angular-momentum-observer orbit, suggesting that modifications in the 
Casimir energy do appear when the azimuthal symmetry is broken by the equatorial circular motion 
of the cavity with respect to the local zero-angular-momentum-observer frame. A similar analysis has been carried on taking into 
account a Casimir apparatus orbiting an Ellis wormhole \cite{Sorge:ref_1}. While obtaining basically 
the same results, e.g., a decrease in the absolute value of the vacuum energy density inside the cavity, the above work has shown that the Casimir energy is affected by the orbital motion of the cavity nearby the wormhole, also when other local classical measurements 
- performed by a co-moving observer - as the acceleration or the Fermi rotation coefficients 
- yield a vanishing result. In that respect, it is believed that the Casimir effect might represent an 
interesting quantum probe for testing the (non-)inertial character of a given reference frame.

%%%%%%%%%%%%%%%%%%%%%%%%%%%%%%%%%%%%%%%%%%%%%%%%%%%%%%%%%%%%%
\begin{figure}[!]
\begin{center}
\includegraphics[angle=0,width=.49\textwidth]{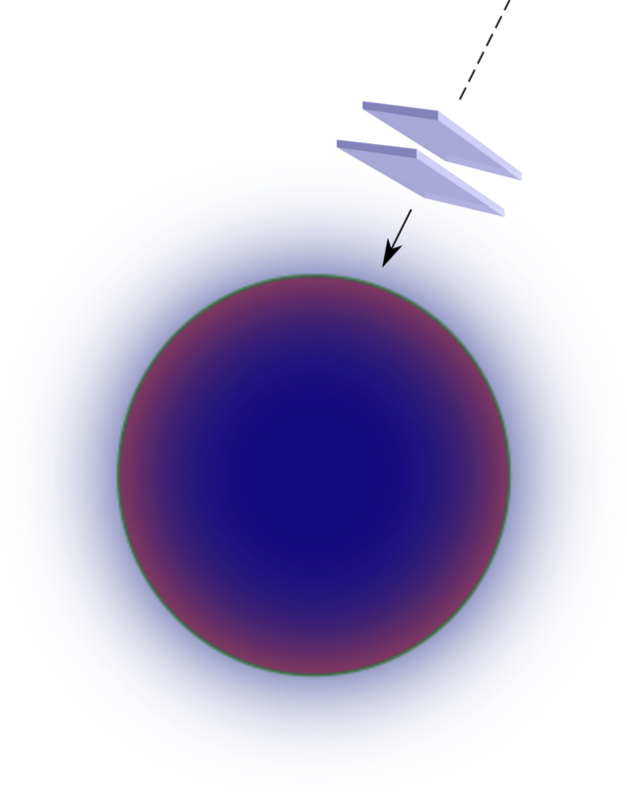}%
\caption{A Casimir cavity freely falling from infinity onto a Schwarzschild black hole. The cavity size is assumed to be small with respect to the Schwarzschild radius, ($L/r_g\ll1$). The cavity is {\em rigid}, i.e., the plate separation $L$ is {\em constant} according to a comoving observer.}.
\label{fig_sorge}
\end{center}
\end{figure}
%%%%%%%%%%%%%%%%%%%%%%%%%%%%%%%%%%%%%%%%%%%%%%%%%%%%%%%%%%%%%%%%%%%

Concerning time-dependent backgrounds, particle creation out of quantum 
vacuum has been obtained by analyzing a ``kicked" Casimir cavity \cite{Sorge:2006pd} and, more recently, a Casimir cavity in free fall into a black hole \cite{Sorge:2019ecb, Sorge:ref_2}. Such effect can be considered as a time-dependent 
tidal one, due to the non-uniformity of the gravitational field experienced by the quantum 
field confined to the cavity, as the black hole horizon is approached. In figure (\ref{fig_sorge}) a schematic picture of a small Casimir cavity radially falling onto a Schwarzschild black hole is shown. The cavity is rigid, namely the plate separation $L$ is constant with respect to a comoving observer. 

Similar particle creation effects can be found when the confined quantum field experiences a 
time-dependent background describing a (weak) gravitational wave. In an early related study on a small three-dimensional cavity \cite{Sorge:2000xk} hosting a scalar quantum field, 
 the relevance of confinement from the point of view of a particle-creation effect has been highlighted. Last, within that framework the excitation of acoustic phonons out of the vacuum in a toy-model one-dimensional elastic medium has been considered, as a consequence of the 
interaction with a gravitational wave \cite{Sorge:2018lhw}. 

Future investigations will be devoted to a deeper understanding of the interplay between gravity and quantum fields both in the static and in the dynamical regime, the latter being typically 
related to particle creation. Moreover, mesoscopic quantum systems 
interacting with a gravitational wave \cite{Sorge:2010zz} will be considered in detail. Quantum effects induced by gravitational 
waves at nanoscales might represent a new stimulating research arena, both from a 
theoretical and the experimental point of view.

Finally, quantum vacuum will be considered at cosmological level, being strongly related to the cosmological constant problem in non-trivial spacetime topologies as well as in models with compact extra dimensions. Let us stress that, along these lines, Gamma-Ray-Bursts \cite{carlson_swanson} - will be investigated, as they could be deeply rooted in the intriguing interplay between GR and QFT and their capability to be indicators of the Universe expansion will be put in question.

\vskip 1.2cm

\section{Tomographic description of quantum and classical states of the universe}\label{cosimo}

\vskip 0.6cm
 
Modern cosmology was born after the formulation of general relativity by Albert 
Einstein. The evolution of the universe is described by solutions of the Einstein equations. 
However, Hawking-Penrose theorems show that these solutions inescapably present singularities if the energy-momentum tensor has properties which are regular when ordinary matter and 
radiation are considered. The existence of these singularities, which appear to be a pathological aspect of the 
theory, can be interpreted as a signal that the theory is no longer reliable at particular 
conditions, for this reason many researchers postulated that general relativity, which is a classical theory, 
must be replaced by a quantum theory of gravitation. In particular, cosmology seems a topic 
that lends itself well to the study of the quantum properties of a gravitational system such as 
the universe, which is homogeneous to a very high degree; it is therefore sufficient 
to study the problem of quantization in the minisuperspace, which is the space of 
all homogeneous three-metrics.

The study of the evolution of the universe is the story of the evolution of the 
states of the universe. The state of a particle in classical physics is described by its 
position and momentum in phase space. More generally, for a particle in a thermal bath, 
its state can be described by a probability function. For a constrained system where not all the dynamical variables are independent (for example the Friedmann equations show that expansion factor $a(t)$ and its time derivative $\dot{a}(t)$ and are not independent), it is sufficient 
to use a smaller number of variables, and the states are confined in a subset of phase space. 
Given one state and the equation of motion, one can predict the evolution of the physical system.

On the other side, in quantum theory the state of a particle, or more generally a quantum system, is defined by a wave function (or functional 
in field theory). It can be used to evaluate 
the probability that a physical system is in a particular 
state and to obtain the probability amplitudes of its transition from a state to another. 
Alternatively to the wave function, other descriptions of the physical state have been introduced 
like the density matrix or the Wigner function. The last one describes the physical state in a
phase space, even if, differently from the classical particle, it cannot be considered as a probability 
function since it can take negative values.

In general relativity the existence of singularities does not make it possible to predict
the evolution of the universe from an initial undetermined state. On the other side the quantum description of the initial states of the universe needs to be extended to the final states crossed by the classical universe. Unfortunately, there is no a simple way to perform
this extension by using the kind of representations used so far. For this reason, we introduce a new description in terms of tomograms which are marginal 
probability functions, and as such they are positive-norm functions and are observables (see for example\cite{Manko:1999ydp} and \cite{Ibort:2009bk})
They describe equally well quantum and classical states. For this reason, tomograms are suitable 
for a good description of the quantum-to-classical transition of the universe, and eventually can 
be defined phenomenologically by cosmological observations, leading to a reconstruction 
of the early universe \cite{Manko:2003dp} \cite{Man'ko:2004zj}
\cite{Man'ko:2006wv}
\cite{Capozziello:2007xn}
\cite{Capozziello:2009ah}
\cite{Stornaiolo:2014hpa}
\cite{Stornaiolo:2015goa}
with obvious relations with its dark components in matter and in energy.

For this purpose, recently a de Sitter universe was considered and the wave functions, 
solutions of the cosmological quantum equation (the Wheeler-DeWitt equation) and were
{\textsl{translated}} in the tomographic formalism \cite{Stornaiolo:2018lvp}\cite{Stornaiolo:2019wau}.

\begin{figure}
\begin{center}
\includegraphics[width=0.7\textwidth]{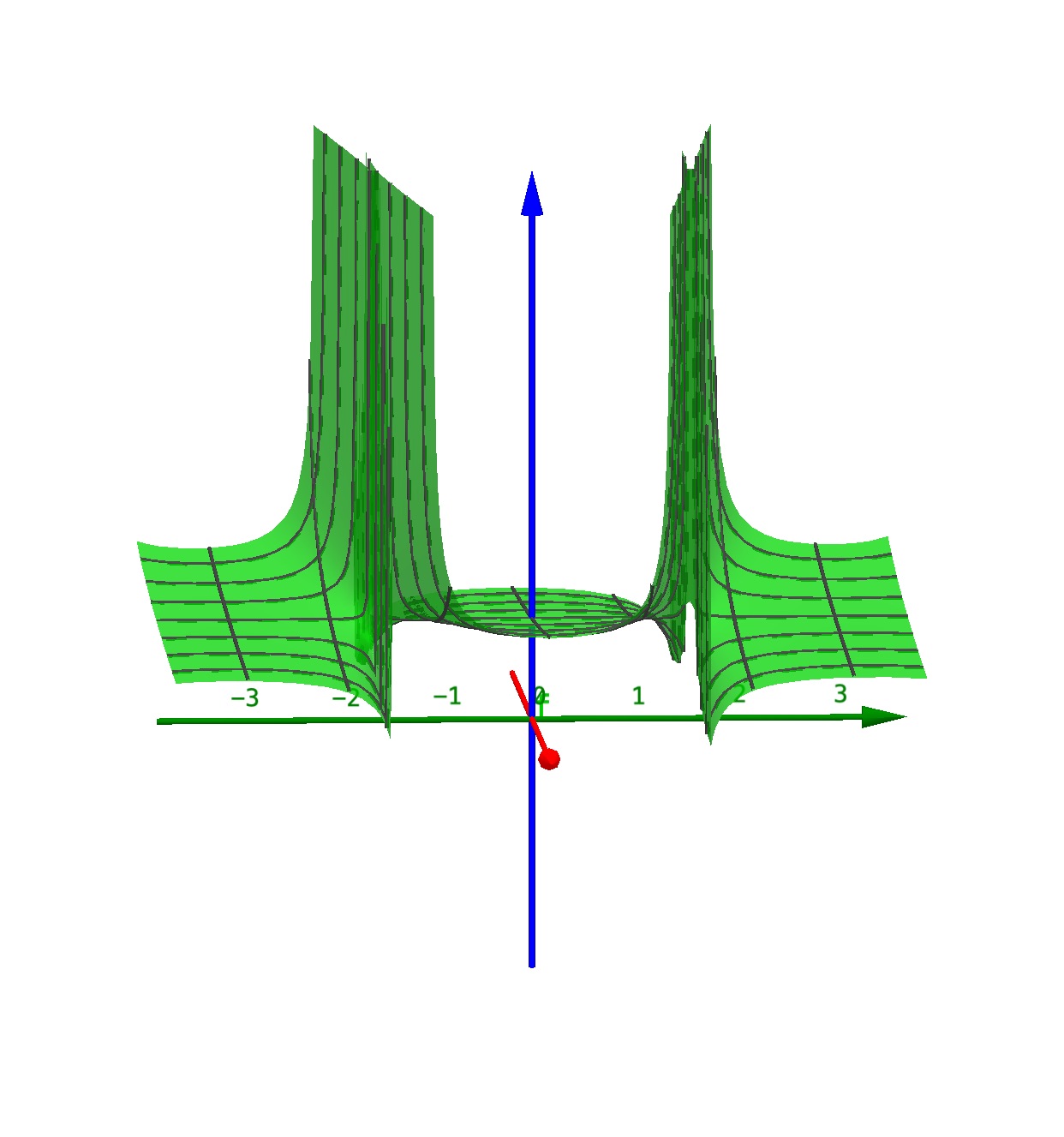}
\caption{The tomogram of a classical de Sitter universe}
\label{classicaltomo}
\end{center}
\end{figure}

If we describe the de Sitter space-time with the closed metric
\begin{equation}\label{metric}
ds^{2}=-\frac{N^2}{q}d\tau^{2}+\frac{q}{1-R^{2}}dr^{2}+qr^{2}(d\theta^2 +\sin^2 \theta d\phi^2)
\end{equation}
where $q=a^2$, $N$ is the lapse function (related to the freedom of choice of the time variable). The classical distribution is represented by the delta function of the constraint equation 
\begin{equation}\label{classicaluniverseconstraint}
f(q,p)=\delta\left( -4p^{2}+\lambda q-1\right) 
\end{equation}
where $p=-\dot{q}/{4N}$ is the conjugate momentum, and the corresponding classical tomogram is (see fig.(\ref{classicaltomo})
 \begin{equation}\label{classicaluniversedetomogram}
 	\mathcal{W}\left(X,\mu,\nu \right)=
	\int\delta\left( -4p^{2}+\lambda q-1\right)\delta(X-\mu q-\nu p)dq dp
	 =\frac{1}{2|\mu|}\frac{1}{\left| \sqrt{\frac{\lambda^{2}\nu^{2}}{16\mu^{2}}+\frac{\lambda X}{\mu}-1}\right|}\, .
\end{equation}

\begin{figure}
\begin{center}
\includegraphics[width=1\textwidth]{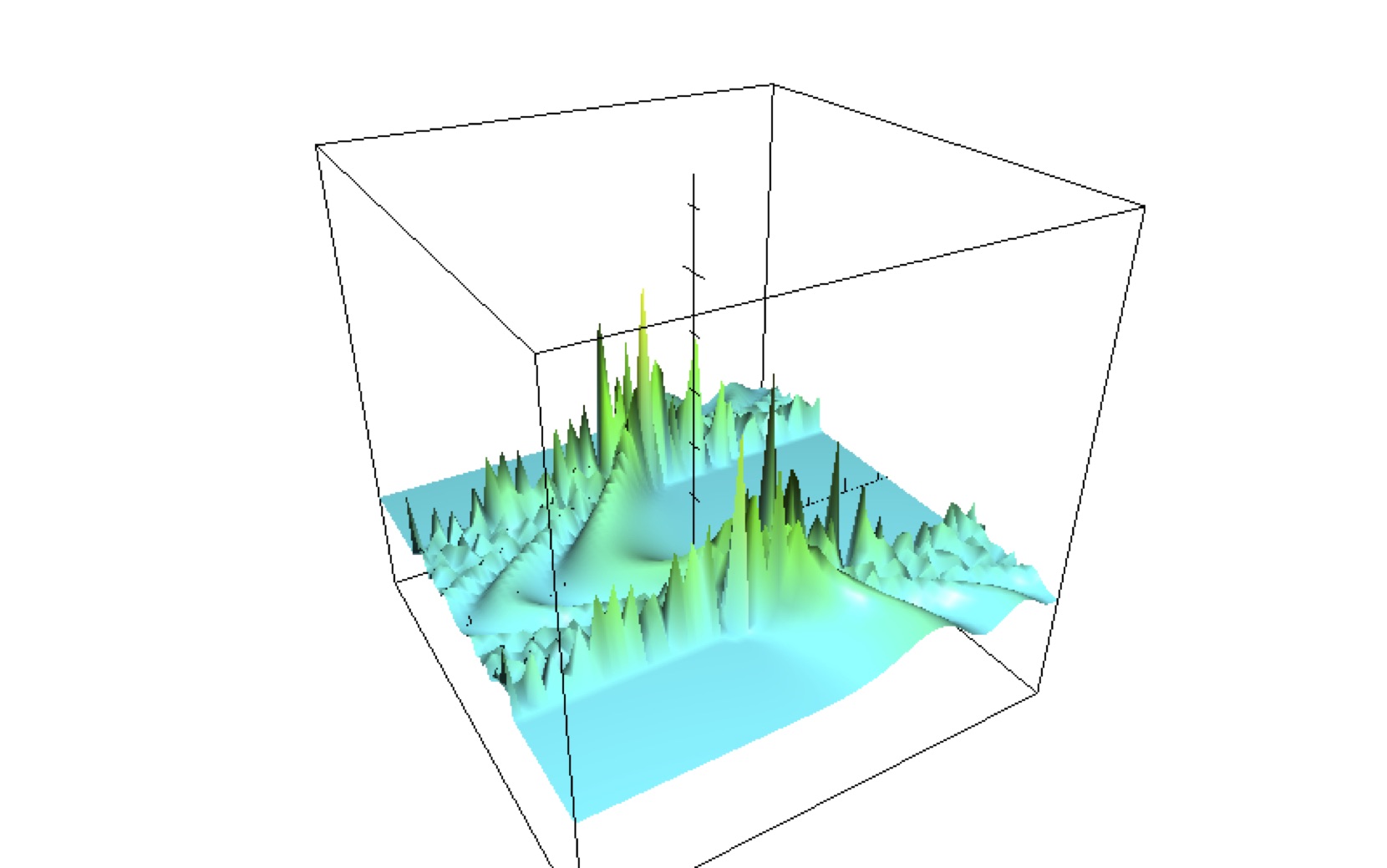}
\caption{The tomogram of a quantum de Sitter universe according to the Hartle and Hawking initial conditions}
\label{HHtomo}
\end{center}
\end{figure}
 
The tomogram corresponding to the Hartle-Hawking wave function is (see also fig.\ref{HHtomo})
\begin{equation}\label{tomo_hh}
\mathcal{W}(X,\mu,\nu)= \frac{A^{2}}{|\mu|}\left|Ai\left( \frac{1}{(2\hbar\lambda)^{2/3}}\left( 1-\frac{\lambda X}{\mu}-\frac{\lambda^{2}}{16}\frac{\nu^{2}}{\mu^{2}}\right) \right) \right|^{2} 
\end{equation}
where the constant $ A =2^{5/6}\pi/\hbar^{1/6}\lambda^{2/3}$.
 These two tomograms can be compared. Then,
first we see that the tomogram resulting from taking the limit for $ \hbar\to 0 $ or the limit for $ \lambda\to 0 $ is the same. This circumstance suggests that a decay of the cosmological constant from the Planck scale to its present 
value causes a transition from the quantum to the classical regime, as the quantum tomogram becomes 
 \begin{equation}\label{negativeAiLimit}
 \mathcal{W}(X,\mu,\nu)\approx \frac{1}{2|\mu|}\frac{1}{\left|1-\frac{\lambda X}{\mu}-\frac{\lambda^{2}}{16}\frac{\nu^{2}}{\mu^{2}}\right|^{1/2}}
 \left| \cos\left( \frac{2}{3}S^{3/2}-\frac{\pi}{4} \right) \right|^{2}\, , 
 \end{equation} 
The Hartle-Hawking model does not converge to the classical tomogram and does not converge at all. But if nevertheless instead of taking $ \lambda\to 0 $, we take $ \lambda\sim 10^{-122} $
we have the possibility of examining a new ``classical" model of universe with peculiar 
features characterized by the presence of the cosine factor in (\ref{negativeAiLimit}). Therefore, the Hartle-Hawking model is not ruled out, but its asymptotic expression is a possible "classical" description of the present universe not derived from the classical equations. In other words quantum cosmology offers the possibility of formulating new classical models that are not necessarily derived from the classical general theory of relativity.

\begin{figure}[h]
\begin{center}
\includegraphics[width=0.6\textwidth]{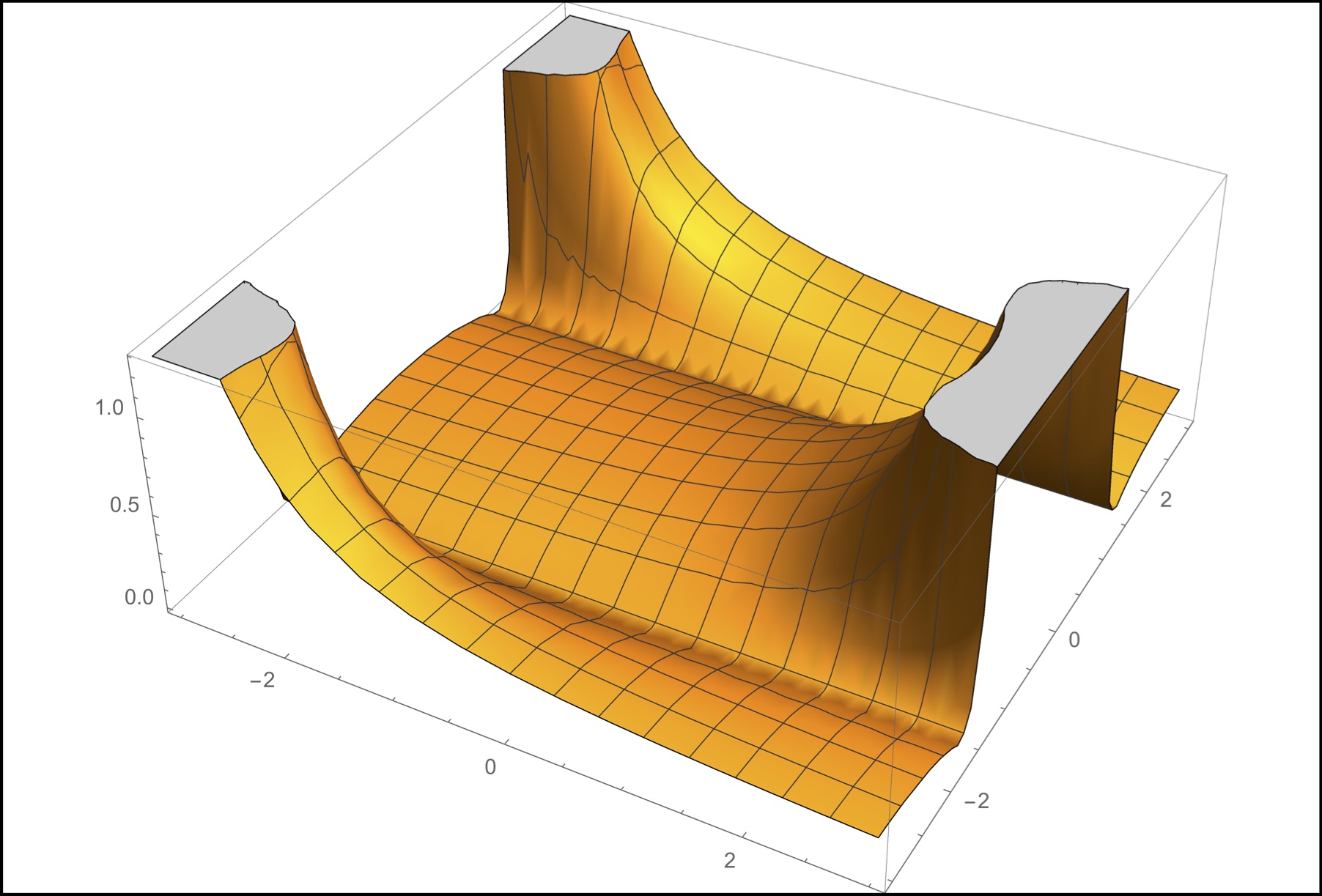}
\caption{The tomogram of a quantum de Sitter universe according to Vilenkin's "tunneling from nothing" initial conditions}
\label{vilenkintomo}
\end{center}
\end{figure}

On the other side the tomogram 
 \begin{align}\label{possiblesolution}
 \nonumber \mathcal{W}(X,\mu,\nu)&= \frac{A^{2}}{|\mu|}\left|\frac{1}{2}\left( Ai\left( \frac{1}{(2\hbar\lambda)^{2/3}}\left( 1-\frac{\lambda X}{\mu}-\frac{\lambda^{2}}{16}\frac{\nu^{2}}{\mu^{2}}\right) \right) \right. \right. \\
&\left. \left. +i Bi\left( \frac{1}{(2\hbar\lambda)^{2/3}}\left( 1-\frac{\lambda X}{\mu}-\frac{\lambda^{2}}{16}\frac{\nu^{2}}{\mu^{2}}\right) \right)\right) \right|^{2}, \end{align}
which is the tomogram obtained from Vilenkin's initial conditions (see fig.\ref{vilenkintomo}) \cite{Stornaiolo:2020vxu}, in the limit $( 2\hbar\lambda)^{2/3}\to 0 $ converges asymptotically to 
 \begin{equation}\label{finaltomogram}
 \mathcal{W}(X,\mu,\nu)\approx \frac{1}{2|\mu|}\frac{1}{\left|1-\frac{\lambda X}{\mu}-\frac{\lambda^{2}}{16}\frac{\nu^{2}}{\mu^{2}}\right|^{1/2}}\,. 
 \end{equation}

 Remarkably, the quantum to classical transition found here can be extended to more general cosmological models involving either scalar fields or cosmological fluids \cite{Stornaiolo:2020vxu}. Many of these models predict that a classical inflationary universe emerges after the decay of the cosmological constant problem during the quantum epoch, and a further decay of the cosmological constant guarantees the exit from the inflationary epoch. A result of this analysis is that the so called "cosmological constant problem" can be addressed in quantum epoch rather than in the classical one.

Last, tomograms are in principle observables. The form of the classical tomogram (\ref{classicaluniverseconstraint}) suggests that the phenomenological tomogram of the universe should be a function of the cosmographic parameters extending the analysis in sect. \ref{Sec:Cosmography} to the variables of the tomogram.

\vskip 1.2cm

\section{On the origin of the cosmological constant: Planckian fluctuations, decoherence scale}
\label{Viaggiu}

\vskip 0.6cm
 
One of the biggest problems in modern physics and cosmology is the nature and origin of dark energy. There 
exist many attempts to explain the current acceleration phase of our universe, by 
adopting `modified gravity theories' or introducing ad hoc running (in time) 
scalar fields. In the standard cosmological model dark energy is interpreted as 
vacuum energy and it is simply depicted in terms of a 
cosmological constant $\Lambda$, both in time and space. The observed value for $\Lambda$ is very small and in complete disagreement with theoretical 
predictions (about $122$ orders of magnitude smaller than the value predicted by Quantum Field Theory). 
Scarsely convincing explanations concerning the issues related to $\Lambda$ have been put forward (see for example \cite{ref_v3}, \cite{ref_v4}, \cite{ref_v5}, \cite{ref_v6} and \cite{ref_v7} and
references therein). It is important to point out also that all the above bounces on the other great darkness of the Universe: the dark matter component, whose interaction with the dark energy is as important as presently unknown.

Remarkably, a completely new point of view, 
regarding the origin of $\Lambda$ has been recently put forward \cite{ref_1}, \cite{ref_2} and \cite{ref_3}
The main idea relies on the nature of $\Lambda$
and its very small observed value and states that quantum fluctuations can explain the equation of state of 
the cosmological constant. In fact, a radiation field sufficiently 
close to the Planck scale can allow quantum fluctuations so strong to permit the 
transformation of this radiation field into one
with the equation of state that $\Lambda$ shows. The effective value of the cosmological constant depends on the physical scale under consideration. 
Thus, in the new view, Planckian fluctuations average on bigger and bigger scales: the
cosmological constant is watered down on scales bigger than the Planck one. The important concept of decoherence scale $L_D$ is also introduced:
this is the scale determining the observed value of the cosmological constant 
and determines the crossover to classicality. The calculations concerning the formula of the observed cosmological constant $\overline{\Lambda}$ in terms of the decoherence scale $L_D$ are performed
in the context of a semi-classical solution where the effective
metric at the Planck scale is averaged. In formulas we have:
\begin{equation}
\overline{\Lambda}=\frac{\pi^3}{90}\frac{L_P^2}{L_D^4},
\label{30}
\end{equation}
where $L_P$ is the Planck length and the constant
$\frac{\pi^3}{90}$ has been fixed in \cite{ref_2} by analogy with Casimir effect. 
From a mathematical point of view the Buchert formalism in
\cite{ref_v8},
known since 2001 and used to model inhomogeneities in our universe by means of a template homogeneous metric,
is properly modified and applied at Planckian scales. Also this research field represents a novelty and, in absence of a sound quantum gravity proposal, provides a practical way to perform 
physically sound calculations.

A comparison with the Casimir effect is outlined in \cite{ref_2}. 
In this regard, it should be noticed that an expression for a quantum modification of the Misner-Sharp mass, introduced in
\cite{ref_v9}, can be written in the following way:
\begin{equation}
E(L)=\frac{c^4}{2G}\frac{L^3}{L_A^2}+\xi\frac{c^4}{2G}\frac{L_P^2}{L}.
\label{11}
\end{equation}
In this context, as usual, 
the Casimir energy $E_C$ can be expressed as the difference between vacuum energy in presence of boundary conditions $E_B$ and the one of a space $E_0$ without boundary conditions:
\begin{equation}
E_C=E_B-E_0.
\label{23}
\end{equation}
and as a result one obtains the expression $$E_C=E(L,\overline{\Lambda})-E(L,\Lambda)=\frac{\xi c^4}{2G}\frac{L_P^2}{L}$$,
in perfect agreement with the Casimir energy $E_C$ obtained in the standard Casimir effect for a spherical conductor with $\xi=\frac{\pi^3}{360}$.
This comparison allows to fix the 
decoherence scale $L_D$ to be of the order of about $10^{-2}$ millimeters. Note that this is 
the scale where thermodynamic fluctuations are expected to be of the same order of the 
quantum ones in the standard Casimir effect.
Future studies must be focused in order to depict the decoherence scale in terms of Quantum Field Theory approach. By virtue of the well known lack of a well posed quantum gravity theory, we need an effective treatment of the Einstein's equations at Planckian scales incorporating sound arguments of general relativity and quantum mechanics. The semi-classical models so obtained can 
guide us towards a complete and accepted quantum gravity theory. Moreover, extended theories of gravity naturally generalize the Misner-Sharp mass: it is thus intriguing to explore such modifications according to the Casimir effect for spherical configurations.
 
 Finally, in the same line of research, in \cite{ref_4} and \cite{ref_5} the logarithmic corrections to 
Black Hole entropy have been outlined
in a very simple and physically sound way. A similar formula can be applied to the whole universe represented by its apparent horizon, allowing the study of thermodynamics' laws 
at a cosmological scales. Then, in \cite{ref_6} and in \cite{ref_7},
a generalization of the well known Bekenstein-Hawking entropy formula 
at a cosmological level has been obtained: in an expanding universe a further term arises 
depicting the degree of freedom due to the non-static nature of our universe. 
Recently, in \cite{ref_8} Saha showed that with this new proposal, according to recent astrophysical data, the phantom era is ruled out. This intriguing fact places in an even more important position the role of $\overline{\Lambda}$.

One can study a relation between the model above for the cosmological constant and the renormalization group (RG) approach. In particular, one chooses a suitable renormalization scale $\mu$ or a subtraction point, where the physics is calculated and fixed, as suggested by \cite{ref_1},\cite{ref_2} and in \cite{ref_3} in a cosmological context. A physically sound well motivated expression for $\mu$ will represent the starting point to formulate a RG equation for the cosmological constant. In this regard, the main goal is to obtain a non-trivial (non-vanishing) infrared (IR) fixed point at $\mu=\mu_D$, thus identifying the fixed point $\mu_D$ with 
the decoherence scale $L_D$. This result will allow to firmly establish the decoherence scale $L_D$ as an invariant one where the value of the 
cosmological is fixed, thus representing the crossover to classicality. 

The physical scales where quantum fluctuations act in a non-trivial way can be of interest, as an example, also in the fate of a collapsing star. 
In particular, are possible modifications of the Chandrasekhar-mass limit (i.e., the maximum mass of a stable white dwarf star)
leading the the formation of more massive stable stars without forming a black hole?. For this purpose, it should be noticed that a possible presence of gravitational-echoes in the Ligo-Virgo detection \cite{ref_9} of post-merger signals of binary coalescent stars could model some deviations from General Relativity in the near-horizon structures or more generally for ultra-compact stars made of possible exotic matter, where quantum effects are expected to come into action. For example, gravitational-echoes are supposed to be generated by the collision of two black holes and may indicate possible new physics beyond General Relativity.
Detections of gravitational-echoes will be the smoking gun of quantum effects in a very strong gravitational regime. Moreover, these objects may be part of the elusive Dark Matter component of the Universe.

\vskip 1.2cm
 
\section{Gravity meets quantum mechanics: a simple quantum low-energy gravity model at work}
\label{NaddeoAdele}

\vskip 0.6cm

The idea that quantum mechanics (QM) should be characterized by a fundamental non-unitary time evolution of the state vector, where gravity plays a prominent role, has been put forward by a number of authors, on different grounds \cite%
{Penrose,Diosi,Bassi1}. Thus two main questions arise, which require an answer: how to reconcile the basic
processes of QM, \textit{i.e.} the Schr\"{o}dinger deterministic evolution
and the non-unitary process associated with the act of measurement, and how to explain the transition to classicality.

These issues appear to be deeply connected with quantum gravity and, in particular, with the way in which
gravitational fields are generated by quantum matter. It is still widely debated if gravity has to be quantized at all or it is intrinsically
classic and should be properly treated. In the latter case a proposed route is the semi-classical gravity, based on the requirement that the energy-momentum tensor appearing in Einstein equations is taken as an expectation on the quantum state. This prescription, together with the Newtonian limit, leads to the Newton-Schr\"{o}dinger equation \cite{Bassi2}. While its solutions for a lump of matter show interesting properties, this equation cannot be a valid candidate to give self-gravity at low energies because the non-linearity allows for superluminal communications.

The effective low-energy model of Newtonian gravity introduced by S. De Fliippo twenty years ago, termed Non-unitary Newtonian Gravity (NNG) (see e.g. \cite{defilippo2,defilentropy,defilippo1}, references therein), belongs to this line of research. It is obtained as the non-relativistic limit of a non-unitary version of higher derivative gravity, which is classically stable and allows in principle for regularization of gravitational collapse singularities \cite{defilippo1}, leading to results in agreement with Bekenstein-Hawking entropy \cite{bek,haw}. Among its appealing features, besides a non-Markov evolution, one finds that the ordinary Newton's action-at-a-distance interaction is recovered at a macroscopic level. Furthermore it exhibits a a mass threshold for dynamical self-localization, which amounts to $10^{11}$ proton masses for ordinary matter
densities \cite{defilippo2,defilippo1,noi}. As a consequence, even for closed systems, macroscopic coherent superpositions of states evolve into ensembles of pure states thanks to the peculiar dynamics of NNG model. 
Thus, as a further bonus, density matrix emerges as the fundamental description of physical reality, characterized by a non-Markovian evolution even from a pure state. 

Due to the above features the NNG model is among the best candidates to address the controversial and still unsolved issue of the quantum foundations of the Second Law of thermodynamics \cite{th2,th6}. Indeed, as elucidated in \cite{wald}, a microscopic derivation  of the Second Law of thermodynamics for a closed system could be obtained only via a non-unitary quantum evolution and in correspondence of suitable initial conditions. This feature is also shared by the process of formation and evaporation of black holes  \cite{haw1}. In this respect a prominent, unifying role is played by the concept of von Neumann entropy as entanglement entropy with hidden degrees of freedom.

Preliminary steps in demonstrating the ability of NNG to give rise to a gravity-induced relaxation towards thermodynamic equilibrium even for a perfectly isolated system have been already performed in Refs. \cite{noi1} and \cite{noi3}. As a first case study, a system has been considered, built out of two particles trapped in a harmonic potential, interacting via delta-like and gravitational interaction \cite{noi1}. By choosing an eigenstate of the physical Hamiltonian as initial condition, numerical simulations have been carried out in order to study the time evolution of von Neumann entropy. The picture is consistent with the interpretation of von Neumann entropy as an entanglement entropy with some hidden degrees of freedom, which is clearly reminiscent of well known black hole entropy calculations \cite{ent1}. As a further result, entropy fluctuations have been found to take place, thanks to the non-unitary part of gravitational interactions, with the initial pure state evolving into a mixture. 

These promising results have been confirmed by switching to a more involved system, i.e., a harmonic nanocrystal within a cubic geometry \cite{noi3}. In this case a numerical simulation has been carried out as well, but the choice of initial conditions is much more involved. By following the procedure outlined in Refs. \cite{Deutsch,Rigol}, an initial pure state with mean energy $E$ has been selected, built up uniformly at random as a superposition of a huge
number of energy eigenstates within the energy interval $\Delta E$ around $E$. The von Neumann entropy as a function of time exhibits a sharp monotonic increase, followed by a stabilization at late times. These findings clearly show that a micro-canonical ensemble grew up within the initial energy levels, as a consequence of a non-unitary gravity-induced relaxation towards thermal equilibrium.

Now a further step has to be performed in order to confirm the above scenario, which implies the simulation of a real crystal. Of course, it is the fundamental non-unitarity of the NNG model which is responsible of a net entropy growth for the system as a whole. Thus a microscopic derivation of the Second Law of thermodynamics can be self-consistently carried out, which makes the NNG model the first low-energy gravity model leading in a natural way to the emergence of Thermodynamics even in a closed system \cite{Naddeo1}.

A second topical issue, concerning causality violations in Newton-Schr\"{o}dinger equation \cite{Bassi2}, has been fully addressed within the NNG model as well \cite{NotaDeFilippo,noi2}. Here the single particle Newton-Schr\"{o}dinger equation has been obtained as the mean-field approximation of an equation of $N$ identical copies of the particle, interacting each other via gravitational interaction, when $N$ goes to infinity \cite{sergio}. The starting point is a general $N$-copy model, which is a fully consistent quantum theory without causality violation problems thanks to the validity of the no-signaling condition \cite{Gisin,polchinski}. Furthermore, while superluminal communications are suppressed, new strange communications among Everett branches of the wave function \cite{noi2} are shown to appear, in close agreement with previous findings by Polchinski \cite{polchinski}.
Within NNG model the density matrix fully characterizes the state of a system. This may suggest to assume the Everett Many World Interpretation as the most natural conceptual framework of that theory. But, at odds with other non-linear approaches to quantum mechanics, here the possibility of constructing an Everett phone between different branches of the wave function appears to be strongly inhibited. 

The implementation of an Everett phone deserves further investigations and is currently under study \cite{Naddeo2} together with the search for a general proof of the existence of gravity induced correlations between the different branches of the wave function within a path integral formulation of the NNG model a' la Feynman-Vernon \cite{FeynmanVernon}. We strongly believe that this issue will help to clarify the role of gravity induced de-coherence in wave function collapse and stimulate to look for the answer to a further question: how to discriminate in principle this kind of fundamental de-coherence against the usual environmental de-coherence and, finally, how to experimentally detect such a difference. To consider astrophysical consequences of NNG model is premature but one can forecast non- GR violation of Newtonian law at interesting scales.

\vskip 1.2cm

\section{Physics meets Cosmology: Turbulence in SPH simulations of galaxy clusters}

\vskip 0.6cm

The author \cite{va19} presented results from a large set of N-body/SPH hydrodynamical cluster simulations aimed at studying the 
statistical properties of turbulence in the ICM (InterGalactic Cluster Medium). 
The cosmological model is a flat CDM model, with vacuum energy density
$\Omega_\Lambda=0.7$, matter density parameter $\Omega_m =0.3$ and Hubble constant $h=0.7=H_0/(100 \ km \ sec^{-1} \ Mpc^{-1})$. $\Omega_b=0.019 \ h^{-2}$ is the value of 
the baryonic density.

The numerical hydrodynamical scheme employs a SPH formulation
in which gradient errors are strongly reduced by using an integral 
approach \cite{va16}.
The ensemble of hydrodynamical cluster simulations
has been constructed by performing a set of
 individual runs, with initial conditions for each cluster extracted
from a cosmological N-body simulation with only dark matter.
We consider both adiabatic and radiative simulations. 
For the cooling runs, the modeling of the gas incorporates
 radiative cooling, star formation and energy feedback from supernovae.

 A gas particle is eligible to form a star particle if the gas flow is locally convergent and the gas density exceeds a given threshold. If these conditions are
 satisfied, star formation will occur with a characteristic dynamical
time-scale. We refer to \cite{va06} for
a detailed description of the recipes implemented. 

We construct clusters subsamples according to the cluster dynamical status or gas physical modeling. We quantify the cluster dynamical state 
by using, as a morphological indicator, the power ratio method \cite{va19}. 
According to this method, the projected X-ray
surface brightness $\Sigma_X(\vec x)$ is the source term of the pseudo
potential $\Psi(\vec x)$ which satisfies the 2-D Poisson equation.
 A multipole expansion of the solution then gives the moments which are used to
 provide an unambiguous detection of asymmetric structures.

To analyze the turbulent velocity field of the simulated clusters we 
introduce a filtering procedure, aimed at decomposing the fluid velocity 
into a large-scale component and a small-scale part. 
We use an iterative multifilter approach, in which mean velocities are 
estimated locally using an adaptive filtering scheme \cite{vaz12}.
 We then extract from cluster subsamples 
small-scale turbulent velocities, obtained by applying to cluster velocities 
the multiscale filtering method. 
We also quantify the statistical properties of the small-scale 
turbulent velocity field by estimating the velocity power spectrum $E(k)$.

The main results of that work can be summarized as follows \cite{va19}.
The velocity power spectra of non-radiative relaxed clusters are mostly 
solenoidal and exhibit a peak at wavenumbers set by the injection scales 
$\simeq R_{200}/10 $.
 The radius $ R_{\Delta}$ is defined such that
\begin{equation}
 M_{\Delta}= (4 \pi/3) \, \Delta\, \rho_\mathrm{c} \, R_{\Delta}^3
\end{equation}
denotes the mass contained in a sphere of radius $R_{\Delta}$
with mean density $\Delta$ times the critical density
$\rho_\mathrm{c}$.

In the high wavenumber regime, the spectral behavior is  steeper
than in the Kolmogorov case.  Radiative simulations are characterized by a shallower
wavenumber dependency, which is due to  the injection of turbulence at
small scales. This in turn  originates from the interaction of compact cool
gas cores with the ICM.
 This small scale driving source acts in addition to the usual large scale injection
 mechanisms, thus showing that there are  multiple injection scales which operate to
 generate turbulence in  galaxy clusters.
For cluster dynamically at equilibrium, the enstrophy profiles of
cooling simulations show a power-law dependency over a large range of radii,
and a very small turbulent-to-thermal energy ratio.

 For an  highly relaxed cluster we find in the core very low gas velocities,
as seen by {\it Hitomi}. Similarly, the radial profile of the sloshing oscillation period
is found in accord with Fornax observations.
In particular, the associated Froude number $Fr$ satisfies
$Fr \simlt 0.1$ within  $r / R_{200} \simlt 0.1$.  The number $Fr$  is
introduced to assess the  importance of buoyancy forces relative to
to stirring motions, so that  very small values indicate that
 stratification is important.

Our findings suggest that in 
cluster cores ICM turbulence approaches a stratified anisotropic regime, 
with weak stirring motions dominated by gravity buoyancy forces and 
strongly suppressed along the radial direction. 
In accord with similar findings \cite{mo19}, we conclude that turbulent heating 
cannot be considered the
main heating source in cluster cores and thus provide
 a viable solution to solve the so-called cooling flow problem.
The center of relaxed clusters is often characterized by the presence of
cool dense cores with cooling times much shorter than the age
of the Universe. This implies radiative losses which will lead to
an inward motion producing a `cooling flow' \cite{fa94}
and large mass accretion rates. This is
not observed, and some heating sources \cite{Mc12} must be operating
in the cluster cores to regulate the cooling flows.

In a forthcoming work, we are planning to investigate the emergence of the
 cool-core/non-cool-core (CC/NCC) dichotomy in simulations of merging clusters.
 Specifically, we want to investigate the impact of radiative cooling on the
 survival of CC clusters in a merging environment.

\vskip 1.2cm
 
\section{A new challenge for the $\Lambda$CDM scenario: the extremely dense environments around Clusters of Galaxies}
 
\label{sec:covone}

\vskip 0.6cm

Galaxy clusters, the most massive gravitationally bound cosmic structures, are not isolated systems.
Theoretical models within the framework of the $\Lambda$CDM cosmological model predict that they form in the highest-density regions of the underlying dark matter density field \cite{Mo1996}.
This is widely supported by cosmological numerical simulations that show a detailed picture of the so-called cosmic web, with massive galaxy clusters located in the 
highest-density nodes.
The spatial correlation between the dark matter halos (hosting galaxy clusters) and the matter density field is described by the halo bias $b_h$ \cite{Tinker}:
\begin{equation}
b_h \, = \, \frac{P_h (k)}{P_m(k)}
\end{equation}
where $P_h$ and $P_m$ are the power spectrum of the dark matter halo distribution and the dark matter density field, respectively. Both quantities are direct function of the scale $k$.
Hence, dark matter halos are not randomly distributed, but their clustering is enhanced relatively to the general mass distribution. Theoretical models also show that the matter density beyond the virial region 
is strongly correlated to the dark matter halo mass \cite{Tinker}. Recent observational works have confirmed this picture. In particular, Refs. \cite{Johnston} and \cite{covone} have confirmed the predicted relation 
$b_h (M)$ between the halo bias and the mass $M$ of the dark matter halo (the so-called 1-halo term).
These works used the technique known as stacked weak gravitational lensing \cite{Oguri} (see also \cite{fatibene,bozza}).
Weak gravitational lensing, i.e., the coherent and weak distortion of the image of background sources by means of a foreground gravitational field,is a fundamental tool to measure the mass distribution in cosmic structures. However, the typical small signal-to-noise due to the correlated matter distribution 
beyond the galaxy cluster virial radius is not large enough to allow the measurement of the halo bias in individual systems. 
This problem can be addressed by using stacked samples of galaxy clusters, where the signal-to-noise is
enhanced by averaging the weak lensing signal over homogeneous samples, where clusters are grouped according to some observable quantity (such as, e.g., the optical richness). Therefore, stacked weak lensing provides an effective approach to test the statistical predictions of the theoretical formation scenarios,
as the direct investigation of the correlated matter distribution around 
individual systems is hampered by the lack of deep and wide astronomical surveys around massive galaxy clusters.

\begin{figure}
 \centering
 \includegraphics[scale=0.33]{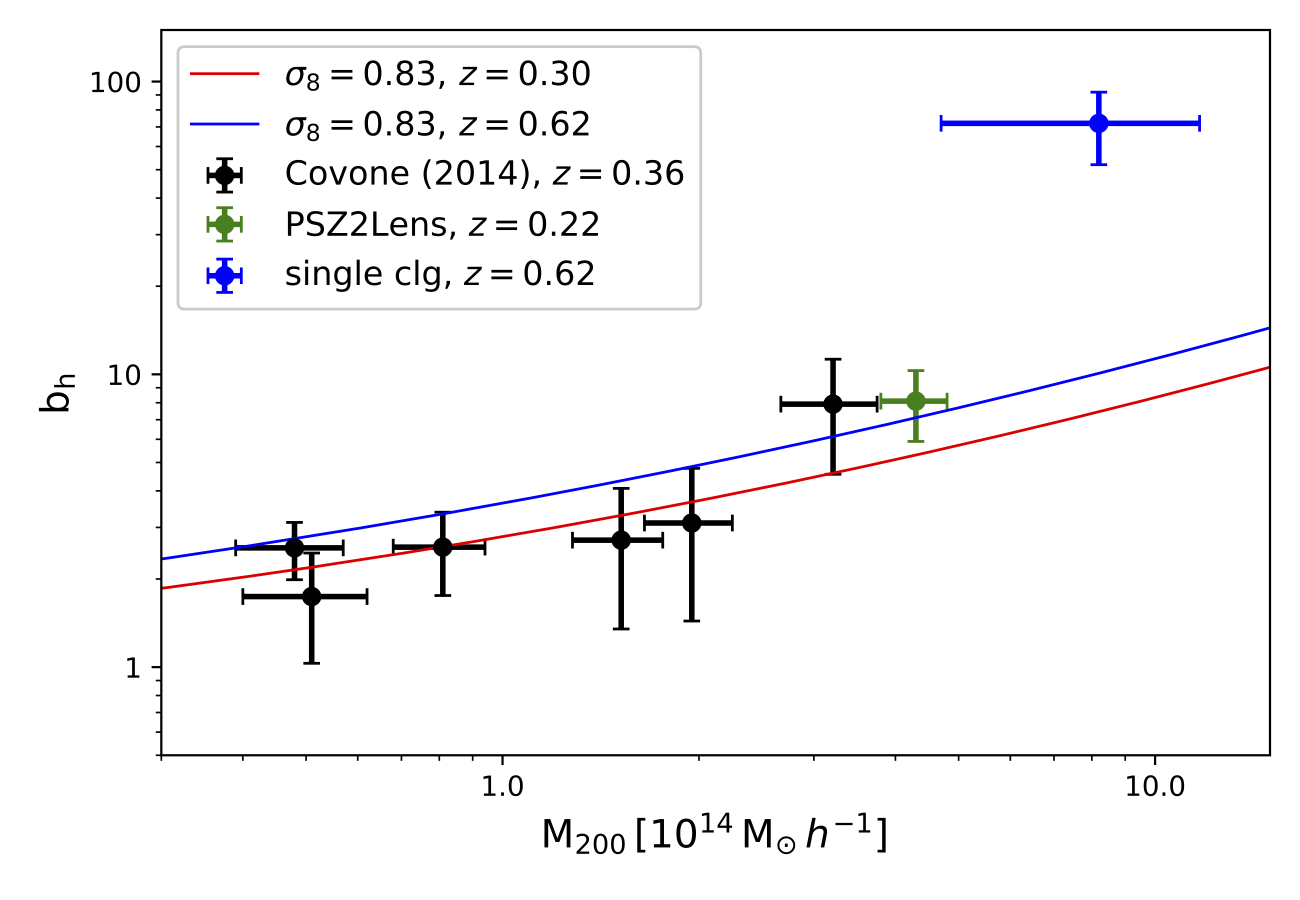}
 \caption{Comparison between the measured values of the halo bias and the theoretical predictions,
 as a function of the virial mass.
 Theoretical predictions are computed assuming a flat $\Lambda$CDM cosmological model with $\sigma_8 = 0.83$. Black points represent the measured values in the stacked sample from Covone et al. (2014) \cite{covone}, at $z=0.36$. The blue point represents the single system at $z\simeq 0.62$.}
 \label{Fig_covone}
\end{figure}

However, the work in \cite{Nature18} recently reported the first detection of 
an extremely dense environment around a massive galaxy cluster, PSZ2 G099.86+58.45, located at redshift $z = 0.616$, from the PSZ2LenS \cite{PSZ2LenS} cluster sample.
PSZ2LenS is a small but homogeneous and complete sample of 35 galaxy clusters, detected by the Planck mission in the sky regions observed by the CFHTLenS (Canada–France–Hawaii Telescope Lensing Survey) \cite{cfhtls} and RCSLenS (Red Cluster Sequence Lensing Survey) \cite{rcs}. Hence
PSZ2LenS is an unbiased sub-sample of the Sunyaev-Zeldovich sources detected by Planck (PSZ2 catalogue).
PSZ2 G099.86+58.45 is a massive cluster (the cluster velocity dispersion is $\sigma_v = 1040 \pm 110 \, {\rm km/s} $) and the most distant system in the PSZ2LenS sample. 
\cite{Nature18} performed the weak lensing analysis by using about $150 \times 10^3 $ background galaxies,
over a sky region about $2.5 \times 2.5 $ square degrees.
Remarkably, PSZ2 G099.86+58.45 is unique in PSZ2LenS, as its large angular diameter distance
($D \simeq 0.98 \, {\rm Gpc} \, h^{-1}$) makes it possible to determine 
the weak lensing signal out to about 30 Mpc from the cluster centre, well beyond the virial radius.
The weak lensing analysis presented in \cite{Nature18} reveals a galaxy cluster mass
in broad agreement with previous measurements based on a dynamical study of the cluster's galaxies motions and on the X-ray temperature profile: $M_{200} = (8.2 \pm 3.5) \, \times 10^{14} M_{\odot} \, h^{-1} .$
However, beyond about 10 Mpc, the environment matter density largely exceeds the cosmological mean. 
Within the $\Lambda$CDM paradigm, the expected halo bias is $b_{h, {\rm \Lambda CDM}} = 11.1 \pm 2.8 $ 
{\cite{Nature18}}. We found a much larger signal from the second halo term: $b_h \, = \, 72 \pm 20 \, ,$
see Fig. \ref{Fig_covone} for a comparison with the theoretical predictions from \cite{Tinker}. 
We note that the measured bias for the whole PSZ2LenS sample is well in agreement with the theoretical predictions, see Fig. \ref{Fig_covone}. 
The observational results obtained by \cite{Nature18} present us with a new dilemma: is PSZ2 G099.86+58.45 a rare case, unexpectedly found in a small (yet complete) sample of galaxy clusters?
Or, is this system the first example of a larger population of clusters whose existence requires a modification of the formation scenario within the $\Lambda CDM$ model? In the latter case, the immediate implication is that enhancing mechanisms around high-mass halos could be much more effective than previously thought.

In the next few years, the aim will be at determining the statistics of the halo bias values around single massive galaxy clusters, by completing a systematic study of the correlated matter distribution around a large sample of galaxy clusters, using recent and future data from wide surveys, such as KiDS and Euclid. 
KiDS (Kilo Degree Survey) is a public survey at the VLT Survey Telescope (Chile) which just completed the observations over about 1500 deg$^2$ in the Southern hemisphere. Euclid, the forthcoming ESA space telescope, will survey about 15 thousand square degrees of the sky, detecting about $10^5$ galaxy clusters up to redshift $z \sim 3$. This wealth of data will allow both to trace the evolution of the correlated matter distribution and to robustly quantify the occurrence of rare systems such as 
PSZ2 G099.86+58.45. The statistics of these peculiar systems will tell us that our Cluster formation theory requires a critical revision.

\vskip 1.2cm

 \section{Merging rates of compact binaries in galaxies and Stellar BHs.}\label{sec:GW}
 
 \vskip 0.6cm
 
 With the recent detections of gravitational waves (GWs) by the LIGO/Virgo collaborations, the era of multimessenger astronomy has begun. The future AdvLIGO/Virgo improved sensitivity configurations, together with the advent of the forthcoming detectors as Einstein Telescope (ET), will provide an enormous number of informations in many different fields: from astrophysics to cosmology and even to fundamental physics (e.g., \cite{taylor+12, barack+19}). In the last years the authors focused on forecasting the merging rates of compact binaries and the associated detection rates with different detectors as a function of redshift \cite{boco+19}. This issue depends on various astrophysical processes happening on different spatial and time scales: we must correctly model single stellar and binary evolution phenomena, GW physics and the environment in which binaries are formed; therefore also galaxy formation and evolution must be kept into account. In the approach of reference \cite{boco+19} the authors have exploited the most recent observations of galaxies' properties (i.e., their UV+far-IR/submillimeter/radio luminosity functions, their spectral energy distribution, the mass metallicity relationships) and combined them with stellar evolution simulations outcomes. This approach can remarkably provide joint probability distribution functions of the host galaxy properties (star formation rate, stellar mass, metallicity, etc.) and the properties of the GW signal.

In order to compute the merging rates as a functions of redshift we rely on three main ingredients: (i) a statistics of galaxies, coming from their UV and IR luminosity functions. 
(ii) a model able to reproduce the observations of the chemical enrichment history of each individual galaxy, (iii) the outcome of stellar and binary evolution simulations.

The first ingredient, given the relation between the luminosity and the star formation rate (SFR) (\cite{kennicutt+99}), can be translated into a star formation rate function at different redshifts; it provides the number density of galaxies producing stars at different star formation rates $\psi$ at various cosmic times/redshifts $\rm t$: $\rm dN/\rm d\psi\,dV$ (see \cite{mancuso+16}, \cite{boco+19}). The second ingredient allows to associate a metallicity to galaxies with different properties (SFR, mass, age or morphological type). The metallicity ($\rm Z$) stellar winds, supernova kicks, direct collapse, common envelope effects all depend on it. Using a very simple model, in agreement with observational data, featuring a rapid linear increase of the metallicity with the galactic age up to a saturation value, the authors of \cite{boco+19} are able to assign a probability distribution function of metallicities at given SFR: $\rm dp/\rm dZ\,(\rm Z|\psi)$ (see \cite{arrigoni+10,gallazzi+14,andrews+13,zahid+14,onodera+16,pantoni+19}). Finally, the outcomes of stellar evolution simulations can provide three crucial factors: the number of merging binaries in an Hubble time per unit of star forming mass ($\rm dN/\rm dM_{\rm SFR}\,(Z)$), the chirp mass probability distribution ($\rm dp/\rm dM_{\rm chirp}\,(\rm M_{\rm chirp}|Z)$) and a time delay distribution between the formation of the binary and the merging ($\rm dp/\rm dt_{\rm delay}\,(\rm t_{\rm delay}|Z)$) (see e.g. \cite{spera+17,dominik+12,belczynsky+16,dominik+15,demink+15,chruslinska+18}).

\begin{figure}[h!]
\centering
\includegraphics[]{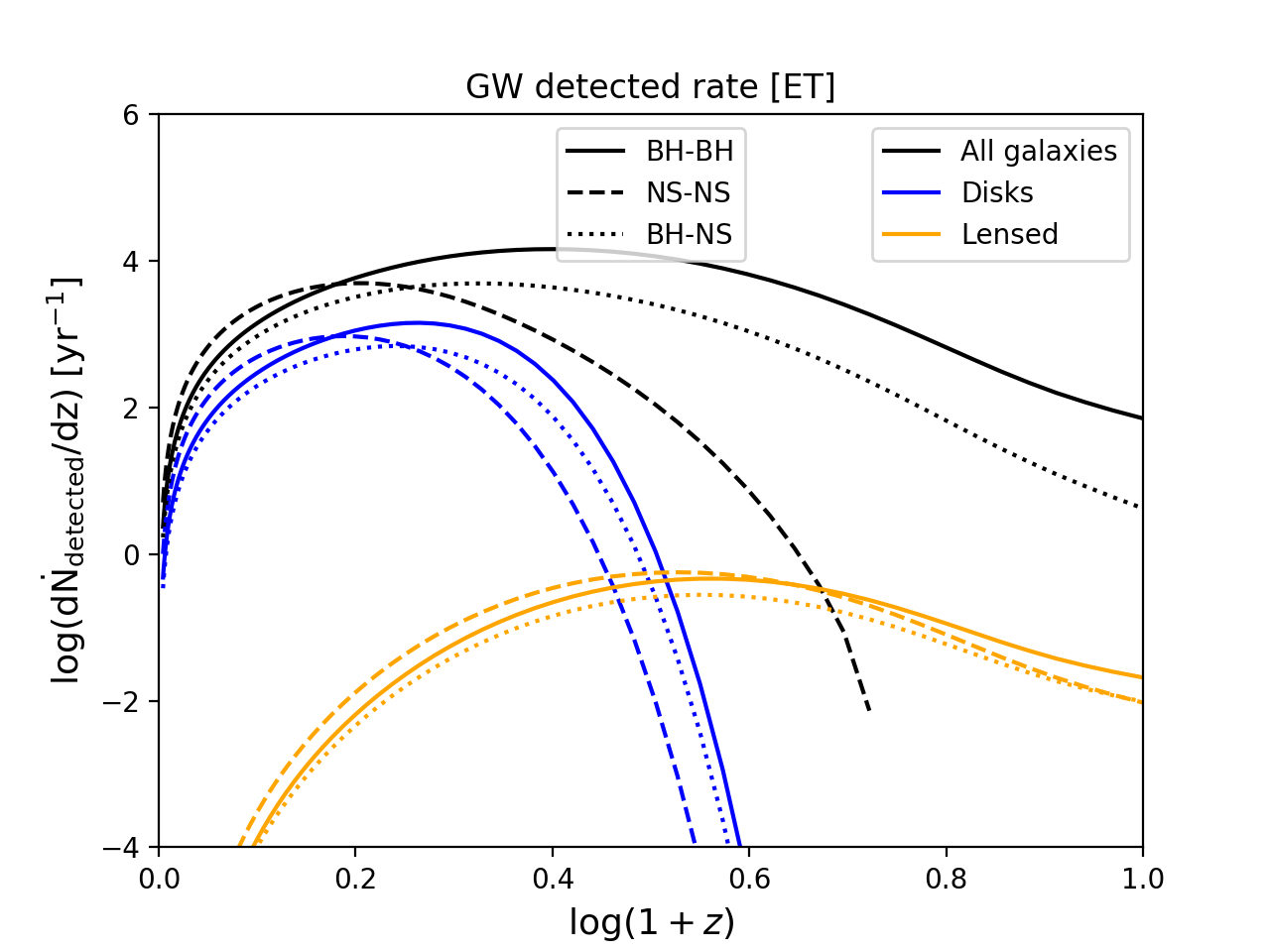}
\caption{GW event rate per unit redshift expected for ET. Solid lines refer to BH-BH events, dashed lines to NS-NS events, and dotted lines to BH-NS events. Black lines refer to the total contribution from all galaxies, while blue lines represent only the contribution of disk galaxies at $z<2$. The orange lines refer to galaxy-scale gravitational lensing of GWs with magnification $\mu>10$.}
\label{fig:detected}
\end{figure}

The ingredients described above can be combined to compute the merging Koushiappasrates per bin of redshift and chirp mass as follows:
\begin{equation}
\begin{split}
\frac{\rm d\dot{N}_{\rm merging}}{\rm dz\,dM_{\rm chirp}}(z,M_{\rm chirp})=&\frac{\rm dV}{\rm dz(1+z)}\int\rm dt_{\rm delay}\int\rm d\psi\psi\frac{\rm dN}{\rm d\psi\, dV}(\psi, z_{t-t_{\rm delay}})\times\\
&\times\int\rm dZ\frac{\rm dp}{\rm dZ}(Z|\psi)\frac{\rm dp}{\rm dt_{\rm delay}}(t_{\rm delay}|Z)\frac{\rm dN}{\rm dM_{\rm SFR}}(Z)\frac{\rm dp}{\rm dM_{\rm chirp}}(M_{\rm chirp}|Z)
\end{split}
\label{merging}
\end{equation}
where $\rm dV/dz$ is the differential comoving volume, dependent on the cosmological parameters, the $1/(1+z)$ factor keeps into account the cosmological time dilation and $z_{t-t_{\rm delay}}$ is the redshift computed at the time of formation of the binary (i.e., the cosmic time at merging minus the delay time $t-t_{\rm delay}$). This equation can be easily convolved with the sensitivity curve of a GW detector (e.g., AdvLIGO/Virgo or Einstein Telescope) to get the detected rate per redshift bin by that detector:
\begin{equation}
\frac{\rm d\dot{N}_{\rm detected}}{\rm dz}(z)=\int\rm dM_{\rm chirp}\frac{\rm d\dot{N}_{\rm merging}}{\rm dzdM_{\rm chirp}}(z,M_{\rm chirp})\int_{\rho_{\rm th}}^\infty\rm d\rho\frac{\rm dp}{\rm d\rho}(\rho|z,M_{\rm chirp})
\label{eq:detected_rate}
\end{equation}
where $\rho$ is the signal to noise ratio (SNR), $\rho_{\rm th}$ is the SNR threshold of detectability and $\rm dp/d\rho$ depends on the properties of the GW signal and on the characteristics of the detector. The result of equation \eqref{eq:detected_rate} is shown in Fig. \ref{fig:detected} for the ET detector.

The approach pursued in \cite{boco+19} allows not only to compute the merging and detection rates, but also to relate the properties of the GW signal (chirp mass or SNR) to the properties of the host galaxy. First of all, since late type galaxies and early type galaxies show a completely different history of star formation and chemical enrichment, by using the approach of \cite{boco+19}, one can understand, depending on the characteristics of the signal, the galaxy type from which a GW event is likely to come. Moreover, one can take equation (\ref{merging}) and not marginalizing over the star formation rate, obtain a joint probability distribution for the chirp mass and the SFR: $\rm d\dot{N}/dz\,dM_{\rm chirp}\,d\psi$. This quantity allows to relate a property directly measured from the GW signal, like the chirp mass, to a property of the host galaxy, i.e. the star formation rate, which can be converted into stellar mass or luminosity. This can result a powerful tool for cosmological investigation through GWs since alleviates the problem of the host galaxy association.

We can also perform tomographic cross correlations between galaxies and gravitational wave signals, as performed by the authors of \cite{scelfo+20}. The outcome of the cross correlations can give useful informations both on astrophysics and cosmology.
Finally the approach described above can be also exploited to compute the number of lensed gravitational wave signals detected by different instruments. Lensed GWs, especially if they have an electromagnetic counterpart, can be a very strong tool to investigate both cosmology and the theory of gravity itself.

\vskip 1.2cm

\section{The Nature of Dark Matter}
\label{sec:DM-nature}
\vskip 0.6cm 

The existence of the "Dark Matter Phenomenon" is fully proven by the presence of a large number of (extremely strong) physical anomalies in the behaviour of both the luminous matter and the radiation. These include, among others: the large-scale structures and the expansion rate of the Universe, the rotational speeds of galaxies (see Fig 14), the weak and strong gravitational lensing of background objects, the extraordinary cosmological object called the  Bullet Cluster \cite{Clowe_2004}, the temperature radial distribution of the hot gas in galaxies and clusters of galaxies and the pattern of anisotropies in the cosmic microwave background (CMB) radiation as detected by Planck. 
Furthermore, the theory of Big Bang nucleosynthesis (BBN), by accurately predicting the  cosmological abundance of the lightest chemical elements, indicates that most of the matter in the universe, so as in every galaxy, cannot be made by baryons \cite{Persic_1992}. 

Remarkably, the dark matter relates with the properties of the entire Universe supporting the case for which a modification of GR cannot explain, {\it alone}, all the above observational evidences.
 %%%%%%%%%%%%%%%%%%%%%%%%%%%%%%%%%%%%%%%%%%%%%%%%%%%%%%%%%%%%%
\begin{figure}[h]
\begin{center}
\includegraphics[angle=0,width=.85\textwidth]{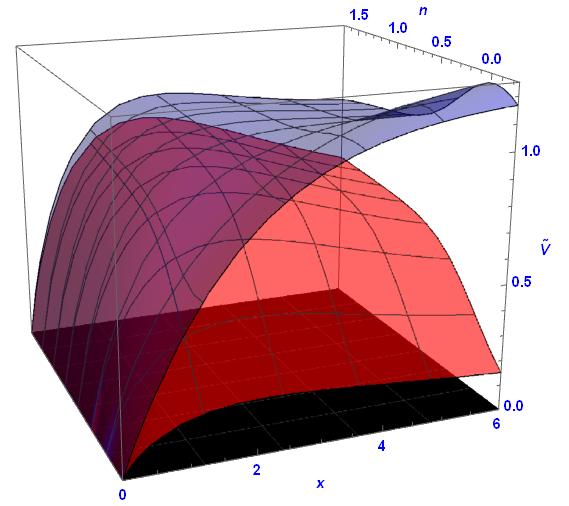}%

\caption{The observed Universal Rotation Curve {\it (blue surface)} of Spirals and its Luminous Matter component {\it (red surface)}. $x\equiv R/R_D$, with $R_D$ the disk lenght-scale, $\tilde V$  is the double normalized URC: $\tilde V\equiv V(x,M_h)/V(3,M_h)$ with $M_h$ the galaxy halo virial mass and: $M_h = 10^{11 + n/5} M_\odot$. The discrepancy implies the presence of a dark massive component (see \cite{PS2007}).}
\label{fig_paolo}
\end{center}
\end{figure}
%%%%%%%%%%%%%%%%%%%%%%%%%%%%%%%%%%%%%%%%%%%%%%%%%%%%%%%%%%%%%%%%%%%

An accurate discussion on various DM particle candidates can be found in (e.g.\cite{Bauer_2017}), here we give a brief account of them. 
It is well known that, for this elusive massive component, there is a favoured scenario: the $\Lambda$-cold dark matter ($\Lambda$CDM) one \citep{Kolb_Turner_1990}; the dark particle is non-relativistic and can be described as a collisionless fluid; it interacts gravitationally with itself and with other massive particles and very weakly with the Standard Model (SM) particles (\cite{Jungman_1996, Bertone}). In detail, the Weakly Interacting Massive Particles (WIMPs) cross section with the SM ones is taken to be that of the Weak interactions: $\sigma \lesssim 3\ 10^{-26}cm^2$; particles in the (1-200) GeV mass range that interact via the electroweak force, with such self-annihilation cross section, imply a relic density of the same order of the observed matter density $\rho_c \Omega_m$. WIMPs are the relic particle from the early Universe, when all particles were in a state of thermal equilibrium. Being the   temperatures of the Universe  $T \gg m_{_{WIMP}}$ the WIMP particles and their antiparticles were both created and annihilated into lighter Standard Model  particles ($DM \ particle  + DM\ particle  \rightleftharpoons SM \ particle + SM \ particle $). As the Universe expanded and cooled down ($ T \lesssim m_{_{WIMP}} $), the average thermal energy of the lighter particles decreased and eventually became too small to form a dark matter particle-antiparticle pair. The annihilation of the dark matter particle-antiparticle pairs did however  continue and the number density of the DM particles started to decrease exponentially ($\propto exp[-m_{_{WIMP}}/T]$) so that the number density became so low that the DM particle-antiparticle interaction vanished. Since then, the number of dark matter particles has remained constant during the continued expansion of the Universe. Because of their large mass, WIMPs move non-relativistically since the Decoupling Time and are candidates of the CDM scenario. Noticeably, they clump  together, from small structures to the largest ones according to the bottom-up scenario.  They have a particular power spectrum of perturbations which guarantees unique initial conditions. It is relevant the fact that Supersymmetric extensions of the standard model of particle physics   have new particles with the above  properties including  the ``WIMP miracle"(e.g. \cite{Steigman_1985, Kolb_Turner_1990}). Crucially, these particle, as any other which is  collisionaless  at galactic scale, lead to DM halos having an  Universal density profile that behaves as: $\rho_{DM} \propto r^{-1}$ in its innermost regions \cite{nfw}.

The Ultralight axion (ULA) with $m \sim 10^{-22} \,eV$ is a scalar field (\cite{Hu_2000, Hui_2017}), that, at large scales mimics the behaviour of a cold particle, while in (small) galaxies, where the inter-particle distance is much smaller than its de Broglie wave length, behaves in a  collective way acquiring  a Bose-Einstein condensates (BEC) equation of state that leads to cored dark halo density configurations, like those observed. 
 
In self-interacting dark matter (SIDM) the DM particles do self-interact with a large scattering cross-section, but with a very small annihilation or dissipation. The relative cross-section is likely due to strong, short-range interactions, similar to neutron-neutron scattering at low-energies, or to weak interactions mediated by the exchange of light particles \cite{Spergel_2000}. Then, the dark matter  particles, inside the originally formed halo cusp, scatter elastically among themselves so that they are  heated up  and leave the region; this effect  transforms the original cuspy density profile into a cored one. The SIDM collision rate is negligible during the early Universe epochs and during the period in which the  cosmological structures form, so the relative cosmological  scenario retains the large-scale successes of that of  $\Lambda$CDM one , affecting the dark structures only at later times and at  small scales. See \cite{Zavala_2013, Kaplinghat_2015}.

The sterile neutrino is a lepton particle beyond the SM of particle physics. (e.g. \cite{Adhikari_2017, Boyarsky_2019}). Its existence is motivated also by arguments on the chirality of fermions and on the possibility to explain in a natural way the small active neutrino masses by means of the seesaw mechanism (e.g. \cite{Asaka_2005}. The mass of this particle when in the keV range (e.g. \cite{Naumov_2019}), being so a warm dark matter (WDM) particle. Created in the early Universe (\cite{Dodelson_1994, Kusenko_2009}) it decouples from the cosmological plasma when it is still mildly relativistic. WDM candidates may account for the various issues at small scales affecting the collisionless CDM scenario; in detail, the fermionic nature of this particle is be cosmologically crucial. In fact, for a $\sim$ keV mass the particle de-Broglie scale length is of the order $\sim$ tens kpc, i.e. the spirals's disks sizes; then,in the latter, a quantum pressure emerges balancing the gravitational force and shaping the inner DM density profile into a cored distribution.(\cite{Destri_2013})The different power spectrum of this particle might account for the observed lack of small mass halos in the local Universe with respect to the outcome of  $\Lambda$ CDM N-Body simulations.  

Recently, there appeared an unexpected new candidate for the Cold Dark Matter scenario that could  raise an amazing connection between the DM in galaxies and the gravitational waves produced by the merging of stellar-mass black holes and detected by LIGO-Virgo experiments. The inferred masses of the merging Black Holes $\sim 50 M_\odot$, infact, seems to be too high to be one of the dead ends of the evolution of the stars in galaxies, and these GW instead may signal the existence primordial black holes created in the very early Universe and working since then as collisionless Macroscopic particles The most interesting case is that in which they provide all the required galactic Dark Matter. Then, a reasonable number of detections of primordial black hole binaries could resolve the nature of this mysterious component \cite{G1}. Let us stress that these objects had continuous merging since recombination and this violent process could have generated a stochastic background of gravitational waves, likely detectable by LISA and PTA. \cite{G2} Of course, just substituting WIMPs with primordial BHs does not immediately relieve the severe tension with the observations at galactic scales that any cold particle has, independently of its mass. 

 The current status of the Universe features the crucial lack of any detection of  Dark particles, expecially the WIMP ones, in a  direct or  indirect way or via super-collider experiment, allied with  the lack of the distinctive central cusp in the DM halo density (e.g.) \cite{cor1,cor2} a set of observed scaling laws among the structure properties of the dark and the luminous matter components in galaxies that are too refined to arise from two of them that just share the same gravitational field (see \cite{sal20}). This challenges the 30-year-old paradigm, that, resting on a priori knowledge of the DM nature, has led us to a quite small number of scenarios led by the collisionless Cold Dark Matter one. Motivated by such observational evidence, it is on the table the idea of resolving the dark matter mystery and its related not understood observations by following a new Paradigm: the nature of DM must be guessed/derived by deeply analyzing the properties of the dark and luminous mass distribution at any scales independently on whether the emerging scenario look to us "main stream" or "exotic". 
 
 An application of this paradigm leads one to propose the existence of a direct interaction between Dark and Standard Model particles which has finely shaped the inner regions of galaxies \cite{Sal20a}. Furthermore, other "exotic" DM candidates, including among many others  e.g. the Mirror Dark Matter (e.g \cite{E1}) and the Strongly Interacting Dark Matter, (e.g. \cite{E2}), neutral dark atoms of composite dark matter {\cite{e3}} appear now reasonable and not anymore crushed by the existence of the WIMP particle a-priori nominated as the actual one. 
 
\vskip 1.2cm

\section{Rotation Curves and Dark Matter in galaxies at high redshifts}
\label{sec:DM-observation}
\vskip 0.6cm
\subsection{Dark halos and the RCs at high z}
\vskip 0.6cm
 We can fairly understand the properties of Dark Matter (DM) in the local Universe (redshift $z \approx 0$) and, by means of the benefits of advancement in the instrumentation, we can now explore the DM properties at $z\sim 1$ to interpret the evolution of Dark and Luminous matter. The latter will certainly contribute to give us crucial information on the nature of DM. 
 In detail, we have the KMOS-Redshift One Spectroscopic Survey (KROSS) data for investigating and understanding the Dark matter scenario at redshift one ($z\sim 1$) in rotation dominated galaxies (disk-type galaxies). KROSS is aimed to study the gas kinematics of redshift one Star-Forming Galaxies (SFGs). Details of the observations can be found in first and foremost papers of KMOS (e.g., \cite{stott2016}, \cite{H17}, \cite{AT2019}). 
 
\cite{GS20} investigated the samples in \cite{H17} (hereafter H17) for the above explained purpose. The sample includes 409 objects with an integrated $H\alpha$ flux $F_{H\alpha}>1.5\times 10^{-17} erg s^{-1} cm^{-2}$, i.e. with an acceptable Signal-to-noise (S/N) and an identified $H\alpha$ emission line. This sample has median redshift $z=0.85^{+0.13}_{-0.04}$, $H\alpha$ luminosity $log(L_{H\alpha}[erg \ s^{-1}])=41.47\pm 0.4$ and effective radii $log(R_e [kpc])=0.45^{+0.23}_{0.33}$. (see TableA1 of H17) Moreover, each galaxy is not an AGN and contains high quality $H\alpha$ spatially resolved data. Effective radii ($R_e$), position angles (PA), inclination angles ($\theta_{i}$), absolute H-band magnitudes ($M_H$), $H\alpha$ Luminosities ($L_{H\alpha}$) and redshifts ($z$) and stellar masses ($M_*$) are also known.
 
 They have modelled the kinematics of the samples under investigation with $^{3D}BAROLO$ code (\cite{ETD15}) that incorporates the instrumental and atmospheric effects (i.e. related to the point spread function (PSF) which determines the spatial resolution and the line spread function (LSF) which corresponds to spatial broadening, in combination these effects are known as beam-smearing.) This is done in a dynamical environment, differently from the 2D maps modelling which is strongly beam-smearing dominated in the inner region of galaxies. By using 3D-Barolo, it is possible to flawlessly compare the data and model in 3D observation space, thus correct for the beam-smearing and other effects dynamically (\cite{ETD15, ETD16}). 
 
 The modelling of 3D-Barolo requires three geometrical parameters, i.e., co-ordinate of galaxy centre in the datacube ($x_0,y_0$), inclination angle ($\theta_i$), position angle (PA) and three kinematic parameters, i.e., redshift ($z$), rotation velocity ($v_{c}$) and dispersion velocity from ionized gas ($\sigma_{H\alpha}$). In their modelling, they precisely define the central co-ordinates of galaxy ($x_0,y_0$), $\theta_i$, $z$ and keep the $v_{c}$, $\sigma_{H\alpha}$ and PA free to fit. This approach allows us to estimate the free parameters in annulii of increasing distance from the galaxy centre without making any assumption on their evolution with radius, yielding a reliable approach on kinematic modelling of 3D datacube. 
 
\begin{figure}[h!]
 \begin{center}
 \includegraphics[width=18.3cm,height=11cm]{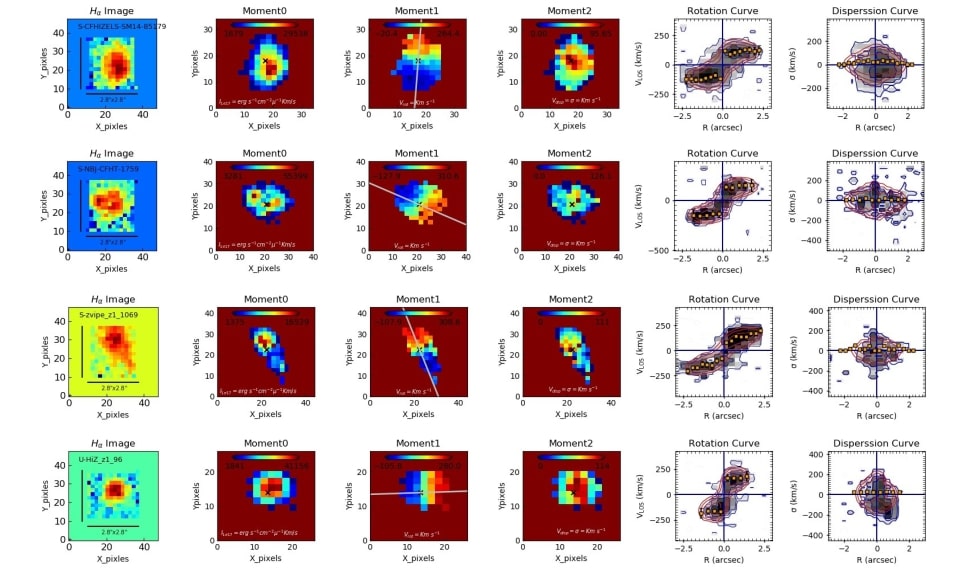}
 \caption{1) $H\alpha$ image from chopped cube; 2) moment0 map; 2) moment1 map; 3) moment2 map; 4) Rotation Curve, black shade is data, red contour is the model and orange square is the best fit $v_{rot}$; 5) Dispersion curve, black shade is data, red contour is model and orange square is best fit, $\sigma_{H\alpha}$ of ionized gas}
 \label{BAROLO-output}
 \end{center}
\end{figure} 
 
Fig. \ref{BAROLO-output} shows a sample of Barolo outputs. From {\it left}  to  {\it right}:  $H\alpha$ image , total $H \alpha$ intensity maps, $H\alpha$ velocity field, $H\alpha$ dispersion field, P-V diagrams in comparison with data and model. Maps are plotted on pixel coordinates (x, y-axis). The black cross shows the kinematic centre of the galaxy, which is in agreement with central photometric coordinates of the galaxy. The grey tilted line is kinematic position angle, red contour in P-V diagram (RC) is model; black shaded area represents the data, and orange squares with error bars are best-fit velocity measurements. 

 \begin{figure}[h!]
 \begin{center}
 \includegraphics[width=17.5cm, height=12.cm]{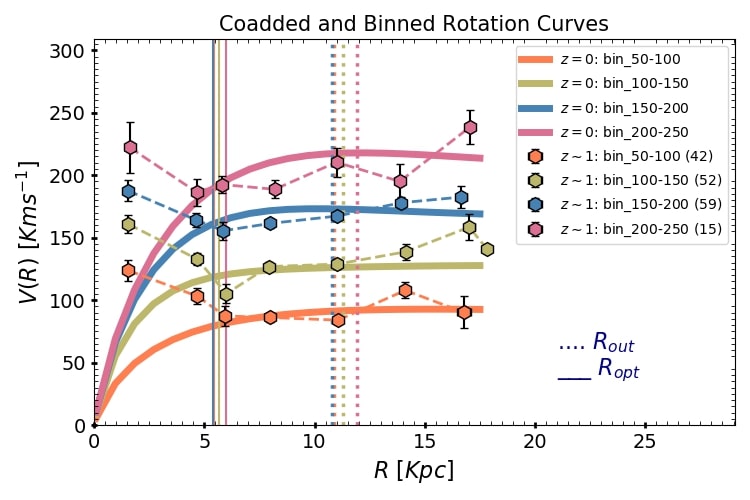}
 \caption{Comparison of $z\sim 1$ (hexagon connected dashed line curves) RCs and those of local spirals derived from the URC (\cite{PS1996, PS2007}) (solid line curves). Colour code is given in the legend. Numbers in the bracket indicate the total number of RCs used in each bin. Dotted and solid vertical lines show  $R_{out}$ and $R_{opt}$ for each RC (colour coded as the velocity bins)}
 \label{RCs1}
 \end{center}
\end{figure}

They have obtained four co-added and binned RCs built out of  from 201 individual RCs. Such a statistical approach has remarkable advantages: 1) it gives us a smooth distribution of RC without being affected by the random fluctuations that arise in each RC from bad data points, i.e., it virtually enhances the S/N in data; 2) it allows mass decomposition of galaxies of similar velocities but having different spatial sampling in the velocity field. This kind of approach in RC studies has been used for decades, pioneered by \cite{PS1991} and  later developed in several  works (\cite{PS1996, PB2000, PS2007, catinella2006, yegorova2011, karukes2017, lapi2018}). In Fig. \ref{RCs1} the results of co-added and binned RCs, in comparison with local RCs are shown. Within $2 Kpc$, there is not data due to the low spatial and kinematical resolution, that affect also the first measure at $2 kpc$. {\it check this} The RCs emerge surprisingly flattish from $R_{opt}=3.2 R_D$ to 4 times $R_{opt}$, and the dark matter starts dominating within $R_{opt}$. The comparison with local RCs shows that in physical units (i.e. $V(R/kpc)/(km s^{-1}) $ the $z\sim 1$ RCs are very similar to the local RCs, though only the mass modelling may clarify this. 

\vskip 0.6cm

\subsection{The Rotation curves profiles and the Nature of Dark matter} 
\vskip 0.6cm

 The careful study of RCs both at low and high redshifts opens the way to pinpoint the nature of the dark particle and to follow the evolution of the DM halos during the galaxy formation era. 

Among other recent works, by studying the RC of a sample of 30 dwarf irregular galaxies, it has been shown that these objects represent new targets for DM indirect searches in gamma rays: in fact, they are DM dominated object with astrophysical gamma-ray emission negligible with respect to the gamma-ray flux expected by DM annihilation events \cite{dIrr}. This claim have been already investigated not only at GeV energy scale, yet for heavy DM candidates at TeV energy scales, e.g. by the High Altitude Water Cherenkov (H.A.W.C) observatory \cite{hawc_irr}. Although (multi-)TeV DM candidates are not favoured by benchmark models in particle physics - e.g. Super Symmetry, after decades of searching no experimental evidence of WIMP DM at GeV energy scale has been found at colliders so far. This fact has generated a crisis in physics and requires an epochal paradigm shift. In view of the new era of TeV observatories the study of particle physics nature of TeV DM candidate is a
well-timed challenge and represents a new frontier in physics.  
 
There are few multi-TeV DM candidates available on a theoretical side. Among others, branons are WIMPs originated by a symmetry breaking mechanism in the extra-dimensional space-time. Thus, for the TeV energy scale new research windows will be opened by the current generation of cosmic-ray observatories (e.g. H.E.S.S.\cite{hess_gc_1, hess_gc_2, bh_gc}, IceCube \cite{ice_bra}) and the next generation of experiments, e.g. Cherenkov Telescope Array (CTA) \cite{cta_bra,cta_GC} and the Square Kilometre Array (SKA) \cite{ska_bra}. These experiments, beside the detection, will able to constraint the WIMPS mass to much higher value than the current one. 

 A careful analysis of high redshift RCs will be crucial for validating any new particle proposed to address the cusp-core issue emerging from all similar analysis of RCs of local galaxies. 
 
One of these is fuzzy DM (FDM). FDM consists of ultra-light bosons that have mass in the range $10^{-23}$--$10^{-20}$ eV,
and arise from symmetry breaking due to the misalignment mechanism in the string theory landscape 
\cite{Hu_2000, 2010PhRvD..81l3530A}. 
Since the de Broglie wavelength of a FDM particle is a few kpc, the density fluctuations on a scale below the de Broglie wavelength are unstable and lead to the formation of a halo whose minimum mass is $\sim 10^7 \, {\rm M}_\odot$ for a boson mass of $\sim 10^{-22}$ eV. Differently, density fluctuations on a scale above the de Broglie wavelength lead to a large scale structure of a DM halo indistinguishable from CDM \cite{2014NatPh..10..496S}. As recently shown in numerical simulations, an imprint of FDM particles is the formation of a core of DM in the innermost part of 
each virialized halo that exhibits a flat density profiles  \cite{2014NatPh..10..496S}. In recent analyses of the kinematic of dwarf spheroidals, ultra-faint and ultra diffuse galaxies favour FDM with boson mass $\sim 10^{-22}$ eV over cuspy NFW profile \cite{2015MNRAS.451.2479M,2019arXiv190210488B,2020arXiv200308313P}. Nevertheless, the debate on the fuzzy nature of DM is far to be closed. The existence of the core in each FDM halo enhances the circular velocity in the inner part of the rotation curve \cite{2019PhRvD..99j3020B}, questioning the model. Additionally, the boson mass constrained with the analysis of the Lyman-$\alpha$ forest data ($ >7\times10^{-21}$ eV) \cite{2016JCAP...08..012B} is almost two order of magnitudes larger than the one required to account for the kinematic of stars in dwarf galaxies ($\sim10^{-22}$ eV). Noticeably, for this particle, high redshift rotation curves will provide us with a decisive test, since in this scenario the DM halos are born with the central density core that we observe today, differently from other scenarios in which the DM cores develop with time from originally cuspy distributions (\cite{Sal20a}). At the same time, SKA and Pulsar Timing Array experiments may play an extremely important role providing us with the smoking gun for FDM by measuring the Compton and de-Broglie scale modulation of Pulsar Timing \cite{2017PhRvL.119v1103D}.

In general, it is very likely once we obtain $\rho_{DM} (r, z)$, where $\rho$ is the DM density , $r $ is the radial coordinate and $z$ the redshift at which this quantity has been measured, we will be able to single out the DM particle. 
 
\vskip 1.2cm

\section{Conclusions}
\vskip 0.6cm

The association of Einstein, Planck and Vera Rubin is made because these great scientists have at least two aspects in common. First, they, as seen together, represent the conjunction point of very different research lines of Physics. Always more people think that such point must be reached in order if we have to succeed in working out that New Physics which is required to frame a currently very large number of unexplained observations and experiment outcomes. The second is that they all provide examples of the time delay, often existing in Physics, between the formulation of a theory or the publication of an intriguing  discovery and that of their general  understanding and validation.

The idea of the great usefulness, for the growth of Physics and our understanding of the Universe, of a contamination between the fields of research of Astrophysics, Cosmology, GR and Elementary Particles is not a recent one but one has to date it 50 years back and recognize it as one of the extraordinary intuitions of Dennis Sciama \cite{s71}. Later on,  other scientists have much  contributed to it: \cite{rev1,rev2,rev3,rev4,rev5} 

Here, we have shown that this idea emerges naturally and roots many sub-fields of Astrophysic, Astroparticle  and Cosmology. These include: the cosmological validation of the beyond GR scalar tensor theories, the (possibly not $\Lambda$CDM) cosmological model for the expansion of the Universe, tensions between different measurements of the same (cosmological model) parameters, the nature of dark matter, the evolution of galaxy halos and their stellar counterparts, the cosmological importance of the stellar BH in galaxies, the evolution of the dark matter perturbations at large scale, the properties of the biggest virialized objects of the universe, the Fundamental Physics related to the scattering angle of two body systems and  in curved space-time backgrounds, the bridge between classical and quantum world, the complex intrinsic neutrino nature and its cosmological importance,  the tomographic description of the Universe, the origin of dark Energy, and the Holy Grail of the coupling between Gravity and Quantum Mechanics. 

Let us stress that this list of fields of investigation covers only a part of all the active fields that also  entangle the GR, the quantum world and the physics of the Universe and that can be successfully  explored  by means of the   multi-lateral approach  we have advocated.  These are,  therefore, subjects for the works of the  present Special Topic. A short list of them includes: the baryogenesis, the  inflation, the formation of primordial and  Supermassive BH, the dark Ages of the Universe, the Astrophysical and cosmological impact of Gravitational Waves.

Finally, it could come as a surprise but, in the above discussion, an unknown entity, that we call dark matter, takes very often the center of the scene. In detail, a large number of issues related to this mystery emerge in many sub-fields of Cosmology, Physics and Elementary Particle Physics. However, one point must be stressed. The investigations of the past decades have not been able to tear the Maja veil off the Dark Matter Phenomenon. Our current knowledge presents mysterious evidences, lacks of congruence and shortcomings in both theoretical and experimental/observational sides. Theories like General Relativity and Standard Model of Elementary particles seem unable to account for the "darkness" of the Universe. As result the answers of our questions about dark Matter do not appear to be straightforward and shining but seem to follow the Nietzsche's directive: Beauty is False, Truth is Ugly. New, multi-lateral approaches to the mysteries of the Universe are likely to be the way to unveil the thick veil above. 
 
\vskip 0.6cm
\section{APPENDIX A: COSMOGRAPHY}
\vskip 0.6cm
\label{App:COSMOGRAPHY}

\textit{Cosmography} is a technique to describe the expansion of the universe without assuming any cosmological model. Such a tool is based only on the cosmological principle, and consists in expanding the scale factor, $a(t)$, in Taylor series around the present time ($t_0$):
\begin{equation}
a(t)=1+\sum_{k=1}^{\infty}\dfrac{1}{k!}\dfrac{d^k a}{dt^k}\bigg | _{t=t_0}(t-t_0)^k\ .
\label{eq:scale factor}
\end{equation}
The different orders of this expansion define the cosmographic series, i.e. the \textit{Hubble}, \textit{deceleration}, \textit{jerk} and \textit{snap} parameters:
\begin{subequations}
\begin{align}
&H(t)\equiv \dfrac{1}{a}\dfrac{da}{dt} \ , \hspace{1cm} q(t)\equiv -\dfrac{1}{aH^2}\dfrac{d^2a}{dt^2}\ , \label{eq:H&q} \\
&j(t) \equiv \dfrac{1}{aH^3}\dfrac{d^3a}{dt^3} \ , \hspace{0.5cm} s(t)\equiv\dfrac{1}{aH^4}\dfrac{d^4a}{dt^4}\ . \label{eq:j&s}
\end{align}
\end{subequations}
 
Noteworthy, the above quantities take on important physical meanings such as the acceleration or deceleration of the universe (by the sign of the parameter $q$), the change of the Universe's dynamics (by the sign of $j$) and the discrimination between an evolving dark energy or a cosmological constant (by the value of $s$).

The Taylor expansion of the luminosity distance can be obtained by the definition $z=a^{-1}-1$ and Eq. (\ref{eq:scale factor}):
\begin{align}
d_L(z)=&\ \dfrac{z}{H_0}\bigg[1+\dfrac{z}{2}(1-q_0) -\dfrac{z^2}{6}\left(1-q_0-3q_0^2+j_0\right)+ \nonumber \\
&\hspace{0.8cm}+\dfrac{z^3}{24}\left(2-2q_0-15q_0^2-15q_0^3+5j_0+10q_0j_0+s_0\right)+\mathcal{O}(z^4)\bigg],
\label{eq:luminosity distance}
\end{align}
which can be fitted by the cosmographic series to trace the expansion of the universe without assuming any \emph{a priori} cosmological model. Indeed, the background evolution can be derived as 
\begin{equation}
H(z)=\left[\dfrac{d}{dz}\left(\dfrac{d_L(z)}{1+z}\right)\right]^{-1}\,,
\label{eq:Hubble rate}
\end{equation}
and find
\begin{equation}
H(z)\simeq H_0\left[1+H^{(1)}z+H^{(2)}\dfrac{z^2}{2}+H^{(3)}\dfrac{z^3}{6}\right],
\end{equation}
where
\begin{equation}
H^{(1)}=1+q_0\ , \quad 
H^{(2)}=j_0-q_0^2\ ,\quad 
H^{(3)}=3q_0^2+3q_0^3-j_0(3+4q_0)-s_0\ .
\end{equation}
\label{eq:Taylor H(z)}
Therefore, it is possible, for a given cosmological model, to calculate the cosmografic coefficients. Furthermore, it is possible to relate the cosmographic series to the cosmological parameters characteristic of a specific model. In the case of the $\Lambda$CDM model:
\begin{equation}
q_0=-1+\dfrac{3}{2}\Omega_m\ , \quad j_0=1\ , \quad s_0=1-\dfrac{9}{2}\Omega_m \ .
\end{equation}
Similarly, for the $w$CDM model, one finds
\begin{align}
&q_0 = \dfrac{1}{2}\left[1+3w(1-\Omega_m)\right], \quad j_0 = 1+ \dfrac{9}{2}w(w+1)(1-\Omega_m)\ ,\\
&s_0 = \dfrac{1}{4}\left[-14-81w(1-\Omega_m)-27w^3(3-4\Omega_m+\Omega_m^2)-9w^2(16-19\Omega_m+3\Omega_m^2)\right], 
\end{align}
and the equation of state of dark energy reads
\begin{equation}
w=-\dfrac{1+2j_0-6q_0}{3(1-2q_0)}\ . 
\end{equation}

\vskip 1.2cm
\section{Appendix B}
\label{App:scatt}

\vskip 0.6cm

We list below the level-5 post-Newtonian-accurate value for the level-4 post-Minkowskian local part of the scattering angle ${\chi_4}_{\rm loc}^{5 \rm PN}(p_\infty)$
\begin{eqnarray}
\label{eq:chi4_loc}
&&{\chi_4}_{\rm loc}^{5 \rm PN}(p_\infty)=\pi \left(-\frac{94899}{32768}\pi^2\nu^2
+\frac{93031}{32768}\pi^2\nu
-\frac{1945583}{33600}\nu+\frac{1937}{16}\nu^2 -\frac{2895}{32}\nu^3\right.\nonumber\\
&& \left.\quad+\frac{525}{64}\nu^4+\frac{1845}{2048}\pi^2\nu^3\right) p_\infty^6 \,.
\end{eqnarray}

\vskip 1.2cm

%%%%%%%%%%%%%%%%%%%%%%%%%%%%%%%%%%%%
\section{Appendix C: Constraining dark energy models at high redshifts}

\vskip 0.6cm
\label{Apphr}
In this Appendix we show the analytical form of the basic cosmological quantities in the different cosmological models investigated in our analisys. For the CPL model we have:

\begin{eqnarray}\label{heos}
 H(z,{\mathrm \theta})_{CPL} &=& H_0 \sqrt{(1-\Omega_m) g(z, {\mathrm \theta})+ \Omega_m (z+1)^3+(z+1)^4 \Omega _{\gamma}{\tilde N}}\, \nonumber 
\end{eqnarray}
 where 
 \begin{eqnarray}
 g(z, {\mathrm \theta})&=\exp^{3 \int_0^z \frac{w(x, {\mathrm \theta})+1}{x+1} dx}\,,
 \end{eqnarray} 
 \begin{equation}
w(z) =w_0 + w_{1} z (1 + z)^{-1} \,,
\label{cpleos}
\end{equation}
\begin{equation}
{\tilde N} = \left( \frac{7}{8}\left(\frac{4}{11}\right)^{\frac{4}{3}} N_{eff}+1\right)\,,
\label{Ntilde}
\end{equation}
and we set $N_{eff}=3$.
\begin{table}[h]
\begin{center}
\scriptsize
\resizebox{13cm}{!}{
\begin{tabular}{cccccccccc}
\hline
~ & \multicolumn{7}{c}{\bf CPL Parametrization} \\
~ & ~ & ~ & ~ & ~ & ~ & ~ & ~ \\
\hline
~ & ~ & ~ & ~ & ~ & ~ & ~ & ~ \\
$Id$ & $\langle x \rangle$ & $\tilde{x}$ & $68\% \ {\rm CL}$ & $\langle x \rangle$ & $\tilde{x}$ & $68\% \ {\rm CL}$ \\
\hline \hline
~ & ~ & ~ & ~ & ~ & ~ & ~ & ~ \\
\hline ~ & \multicolumn{4}{c}{Full dataset} ~ & \multicolumn{4}{c}{No SNIa}
 \\
\hline
~ & ~ & ~ & ~ & ~ & ~ & ~ & ~ \\
$\Omega_m$ &0.23 &0.24& (0.19, 0.27) &0.19 &0.2& (0.16, 0.22) \\
~ & ~ & ~ & ~ & ~ & ~ & ~ & ~ \\
$\Omega_b$ & 0.046& 0.046 & (0.04, 0.047) & 0.055& 0.054 & (0.045, 0.068)\\
~ & ~ & ~ & ~ & ~ & ~ & ~ & ~ \\
$w_0$ &-0.91& -0.92& (-1.1, -0.74) &-0.7& -0.7& (-0.8, -0.62) \\
~ & ~ & ~ & ~ & ~ & ~ & ~ & ~ \\
$w_1$ &-0.8& -0.7 & (-1.05,-0.5) &-0.75&-0.72& (-0.93,-0.48) \\
~ & ~ & ~ & ~ & ~ & ~ & ~ & ~ \\
$h$ &0.69& 0.69 & (0.68, 0.71) &0.67& 0.67 & (0.64, 0.69) \\
~ & ~ & ~ & ~ & ~ & ~ & ~ & ~ \\
\hline
~ & \multicolumn{7}{c}{\bf Scalar field} \\
~ & ~ & ~ & ~ & ~ & ~ & ~ & ~ \\
\hline
~ & ~ & ~ & ~ & ~ & ~ & ~ & ~ \\
$\Omega_b$ & 0.051& 0.051 & (0.049, 0.052) &0.051&0.051&(0.050, 0.0514)\\
~ & ~ & ~ & ~ & ~ & ~ & ~ & ~\\
$\mathcal{ H}_0$ &0.98&0.98&(0.95,0.99)& 0.96& 0.96& (0.94, 0.98)\\
~ & ~ & ~ & ~ & ~ & ~ & ~ & ~ \\
$h$ &0.69&0.68&(0.67,0.695)&0.67& 0.67 & (0.65, 0.68) \\
~ & ~ & ~ & ~ & ~ & ~ & ~ & ~ \\
\hline
~ & \multicolumn{7}{c}{\bf Early Dark Energy} \\
~ & ~ & ~ & ~ & ~ & ~ & ~ & ~ \\
\hline
~ & ~ & ~ & ~ & ~ & ~ & ~ & ~ \\
$\Omega_m$ &0.29 &0.29& (0.27, 0.31) &0.285&0.285&(0.271, 0.298 )&\\
~ & ~ & ~ & ~ & ~ & ~ & ~ & ~ \\
$\Omega_b$ & 0.047& 0.048 & (0.037, 0.052)& 0.045& 0.048 & (0.035, 0.047) \\
~ & ~ & ~ & ~ & ~ & ~ & ~ & ~ \\
$w_0$ &-0.66& -0.67& (-0.85, -0.56) &-0.65&-0.63&(-0.75, -0.53)\\
~ & ~ & ~ & ~ & ~ & ~ & ~ & ~ \\
$\Omega_e$ &0.04& 0.035 & (0.032, 0.043) &0.025& 0.023&(0.009, 0.039)\\
~ & ~ & ~ & ~ & ~ & ~ & ~ & ~ \\
$h$ &0.71& 0.71 & (0.69, 0.71) &0.71&0.71&(0.67,0.73) \\
~ & ~ & ~ & ~ & ~ & ~ & ~ & ~ & ~ & \\
\hline
\end{tabular}}
\end{center}
\caption{Constraints on the parameters characterizing the different dark energy models described above. Columns show the mean $\langle x \rangle$ and median $\tilde{x}$ values and the $68\%$ 
confidence limits.} 
\label{tab}
\end{table}

For the exponential potential we have:
\begin{eqnarray}\label{eq:aHOMtime}
&&a^3(t)={t^2\over 2}[(3\mathcal{ H}_0-2)t^2+4-3\mathcal{ H}_0],\label{eq:aHOMtime1}\\
&&H(t)=
{{2\left(2(3\mathcal{ H}_0-2)t^2+4-3\mathcal{ H}_0\right)}\over{3t\left((3\mathcal{ H}_0-2)t^2+4-3\mathcal{ H}_0\right)}},\label{eq:aHOMtime2}\\
&&\Omega_m={{(4-3\mathcal{ H}_0)\left((3\mathcal{ H}_0-2)t^2+4-3\mathcal{ H}_0\right)}\over[2(3\mathcal{ H}_0-2)t^2+4-3\mathcal{ H}_0]^2},\label{eq:aHOMtime3}\\
&&\Omega_{\varphi}={{(3\mathcal{ H}_0-2)t^2\left(4(3\mathcal{ H}_0-2)t^2+3(4-3\mathcal{ H}_0)\right)}\over
[2(3\mathcal{ H}_0-2)t^2+4-3\mathcal{ H}_0]^2}\,\label{eq:aHOMtime4}\\
&&\varphi(t)=-\sqrt{\frac{2}{3}} \log \left(\frac{6.48}{\left(3 \mathcal{ H}_0-2\right) t^2-3 \mathcal{ H}_0+4}\right)\,,
\end{eqnarray}
where $\mathcal{H}_0$ is a constant. 
$\Omega_{\varphi}$ is the scalar field density parameter.
In order to determine the integration
constants we set the
age of the universe, $t_0$, as a unit of time, $t_0 =1$, so that $a_0 = a(1) = 1$, and $ \mathcal{ H}_0= H(1)$.

For the EDE model we have that:
\begin{equation}
\label{edep}
\Omega_{DE}(z,\Omega_m, \Omega_e, w_0)_{EDE}=\frac{\Omega _e \left(-\left(1-(z+1)^{3 w_0}\right)\right)-\Omega _m+1}{\Omega _m(z+1)^{-3 w_0}-\Omega _m+1}\nonumber
 +\Omega _e \left(1-(z+1)^{3 w_0}\right)\,.
\end{equation} 
Therefore, the Hubble function is:
 \begin{equation}\label{Hede}
H^{2}(z,\Omega_m, \Omega_e, w_0,\Omega _{\gamma}, N_{eff})_{EDE}=\Omega_{DE}(z,\Omega_m, \Omega_e, w_0)+
(z+1)^3\Omega _m + (z+1)^4 \Omega _{\gamma}{\tilde N}\,.
 \end{equation}
 Here ${\tilde N}$ is the same defined in Eq.(\ref{Ntilde}).
 
\vskip 1.2cm
\section{Appendix E: On the inverse $\beta$-decay in the comoving frame}

\vskip 0.6cm
\label{INVB}

This Appendix contains the calculation of the inverse $\beta$-decay
rate in the comoving frame. As remarked in
Sec.~\ref{Luciano}, in this case the process occurs due to the interaction of the proton
with the thermal bath of electrons and anti-neutrinos (Unruh effect).
At tree-level, the following three processes occur (see Fig.~\ref{fig_luciano})~\cite{Matsas:1999jx}:

\begin{equation}
\mathrm{(ii)}\,\,\, p^{+}\,+\,e^{-}\,\rightarrow\, n\,+\,\nu_e\,,\quad
\mathrm{(iii)}\,\,\, p^{+}\,+\,\overline{\nu}_e\,\rightarrow\, n\,+\,e^{+}\,,\quad
\mathrm{(iv)}\,\,\, p^{+}\,+\,e^{-}\,+\,\overline{\nu}_e\,\rightarrow\, n
\end{equation}

Upon quantizing the lepton fields as in ~\cite{Blasone:2018czm}, the evaluation of the decay rate
leads to
\begin{equation}
\label{com}
\Gamma^{(\mathrm{com})}\,\equiv\,\Gamma^{(\mathrm{ii})}\,+\,\Gamma^{(\mathrm{iii})}\,+\,\Gamma^{(\mathrm{iv})}\,=\,\cos^4\theta\, \widetilde{\Gamma}_{1}\,+\,
\sin^4\theta\,\widetilde{\Gamma}_{2}\,+\,\cos^2\theta\sin^2\theta\,\widetilde{\Gamma}_{12}\,. 
\end{equation}
The explicit expression of these three
terms is given in ~\cite{Blasone:2018czm}, 
where it is shown that
\begin{equation}
\widetilde{\Gamma}_1\,=\,\Gamma_1\,,\quad \widetilde{\Gamma}_2\,=\,\Gamma_2\,,\quad \widetilde{\Gamma}_{12}\,=\,\Gamma_{12}\,, 
\end{equation}
with $\Gamma_1$, $\Gamma_2$ and $\Gamma_{12}$
given in Eq.~\eqref{lab}. Then, from
comparison of Eqs.~\eqref{com} and \eqref{lab}, it follows
that the rates in the laboratory
and comoving frames are equal when the flavour representation for asymptotic  states is adopted. 
\vskip 1.2cm
 
\section{Appendix F: Geometric phases for neutrinos propagating in matter}

\vskip 0.6cm

The notion of geometric phase can be defined for any curve in a
Hilbert space $\mathcal{H}$. In the kinematical approach of Mukunda-Simon the geometric
phase of a curve $\Gamma: [s_1,s_2] \rightarrow \mathcal{H}$ is given by

\begin{equation}\label{MukundaSimonPhase}
\Phi_{G}(\Gamma) = \arg \langle \psi(s_{1})| \psi(s_{2} )\rangle
- \Im\int_{s_{1}}^{s_{2}}\langle \psi(s)|\dot{\psi}(s)\rangle d s
\end{equation}
where $| \psi (s) \rangle \equiv \Gamma (s)$. Recalling that the effect of matter on
neutrino oscillations can be included by introducing modified mixing angle $\theta_{m}$
and squared mass difference $\Delta m^2_{m}$, we apply the definition of Eq. \eqref{MukundaSimonPhase} to the propagation of neutrinos in a medium. For the phases associated to a single flavour, we obtain

\begin{equation}\label{SameFlavorPhase}
\Phi^{g}_{\nu_{e}}(t) = \arg \left[ \cos \left(\frac{\Delta m_{m}^{2} t}{4 E}\right) + i \cos 2\theta_{m} \sin \left(\frac{\Delta m_{m}^{2} t}{4 E}\right) \right]
 - \frac{\Delta m_{m}^{2} t}{4 E} \,\cos 2\theta_{m} \ ,
\end{equation}

with $E$ neutrino energy, and $t$ time. On the other hand, for the phases associated to flavour mixing we find: 

\begin{equation}\label{DiffFlavorPhase1}
 \Phi_{\nu_{e}\rightarrow \nu_{\mu}}(t)
= \arg \left[\langle \nu_{e}(0)| \nu_{\mu}(t)\rangle \right] - \Im \int_{0}^{t}
\langle \nu_{e}(t^{\prime})| \dot{\nu}_{\mu}(t^{\prime})\rangle d t^{\prime} \\
= \frac{3\pi}{2} + \phi
+ \left(\frac{\Delta m_{m}^{2}}{4 E}\,\sin 2\theta_{m}\; \cos \phi\; \right) t
\end{equation}

\begin{equation}\label{DiffFlavorPhase2}
 \Phi_{\nu_{\mu}\rightarrow \nu_{e}}(t) = \arg \left[\langle \nu_{\mu}(0)|
\nu_{e}(t)\rangle \right] - \Im \int_{0}^{t} \langle \nu_{\mu}(t^{\prime})|
\dot{\nu}_{e}(t^{\prime})\rangle d t^{\prime} \\
= \frac{3\pi}{2} - \phi
+ \left(\frac{\Delta m_{m}^{2}}{4 E} \sin 2\theta_{m}\; \cos \phi\ \right) t \ .
\end{equation}

\vskip 1.2cm
\section{Appendix G: Rotation Curve Data and Instrument Description}

\vskip 0.6cm

KROSS is an Integrated Filed Spectroscopic (IFS) survey using the KMOS instrument on ESO/VLT. KMOS consists of 24 Integrated Field Units (IFUs); those can place within 7.2 arcminutes diameter field. Each IFU covers the $2.8 \times 2.8$ arcsec in size with $0.2$ arcsec pixels. The targets for the survey are selected from extragalactic deep fields covered by multi-wavelength photometric and spectroscopic data: 1) Extended Chandra Deep Field Survey (E-CDFS: \cite{ECDFS1, ECDFS2}), 2) Cosmic Evolution Survey (COSMOS: \cite{COSMOS}), 3) Ultra-Deep Survey (UKIDSS: \cite{UKIDSS}), 4) SA22 field (see: \cite{SA22}). Some of the targets were selected from CF-HiZELS survey. With these targets the $H\alpha$ emission is shifted into J-band. The median redshift of parent sample (KROSS full data) is $z=0.85^{+0.11}_{-0.04}$. The median J-band seeing of the objects was $0.7$ arcsec, with 92\% of the targets observed within seeing $< 1$ arcsec. The data were reduced using ESOREX/SPARK pipeline (\cite{davies2013}). The end product of the process is a 3D datacube  consisting of 2048 channels. These provide us with  spectra, the line and continuum images and the moment maps (see \cite{stott2016} that are  available at \href{http://astro.dur.ac.uk/KROSS/data.html}{KROSS-website}. 

\vskip 1.2cm
 \section{APPENDIX H: INFN - IS QGSKY}
\vskip 0.6cm 

It is not a coincidence that the authors of this review belong(ed) to the INFN-Iniziativa Specifica QGSKY. The National Institute of Nuclear Physics, INFN, is a main Italian research agency with a portfolio of studies ranging from the infinitely small to the infinitely large. Theoretical, experimental and observational research in subnuclear, nuclear and astroparticle physics are topics of interest for the INFN, which has created and supports special projects, called "Iniziative Specifiche". These connect scientists with different skills and competences to reach the excellence in the scientific results in Fields that are at the forefront of Physics. This IS has a particular attention to Quantum Cosmology and towards the Gravitation and the physics of the Universe both within the framework of standard quantum field theory and General Relativity, and in that of extended theories of Gravity. Most of the researches are focused or end up into the elusive nature of the Dark Matter phenomenon or concern the origin of the accelerated expansion of the Universe. QGSKY is therefore a network devoted at theoretical and observational studies lying at the crossroad among Cosmology, Relativity and Quantum World with a flexibility that allows to tackle this frontier region of knowledge. The recent results obtained inside this IS show very naturally the complex entanglement between Dark Matter and the rest of the Physics, which is also the aim of the current ST.

\end{document}